\UseRawInputEncoding
\documentclass[prl, twocolumn, amsmath, amssymb, superscriptaddress]{revtex4-1}

\usepackage{graphicx, braket, xcolor}

\usepackage[pdftex,colorlinks=true,citecolor=black,urlcolor=black,linkcolor=black]{hyperref}

\newcommand{\+}{^\dagger}
\newcommand{\ha}{\hat{a}}
\newcommand{\nbar}{\bar{n}}

\begin{document}
\title{Bosonic Pair Production and Squeezing for Optical Phase Measurements in Long-Lived Dipoles Coupled to a Cavity}

\author{Bhuvanesh Sundar}
\affiliation{Center for Theory of Quantum Matter, University of Colorado, Boulder, Colorado 80309, USA}
\affiliation{JILA, NIST, Department of Physics, University of Colorado, Boulder, Colorado 80309, USA}
\author{Diego Barberena}
\affiliation{Center for Theory of Quantum Matter, University of Colorado, Boulder, Colorado 80309, USA}
\affiliation{JILA, NIST, Department of Physics, University of Colorado, Boulder, Colorado 80209, USA}
\author{Asier Pi\~neiro Orioli}
\affiliation{Center for Theory of Quantum Matter, University of Colorado, Boulder, Colorado 80209, USA}
\affiliation{JILA, NIST, Department of Physics, University of Colorado, Boulder, Colorado 80209, USA}
\author{Anjun Chu}
\affiliation{Center for Theory of Quantum Matter, University of Colorado, Boulder, Colorado 80209, USA}
\affiliation{JILA, NIST, Department of Physics, University of Colorado, Boulder, Colorado 80209, USA}
\author{James K. Thompson}
\affiliation{JILA, NIST, Department of Physics, University of Colorado, Boulder, Colorado 80209, USA}
\author{Ana Maria Rey}
\affiliation{Center for Theory of Quantum Matter, University of Colorado, Boulder, Colorado 80209, USA}
\affiliation{JILA, NIST, Department of Physics, University of Colorado, Boulder, Colorado 80209, USA}
\author{Robert J. Lewis-Swan}
\affiliation{Homer L. Dodge Department of Physics and Astronomy, The University of Oklahoma, Norman, Oklahoma 73019, USA}
\affiliation{Center for Quantum Research and Technology, The University of Oklahoma, Norman, Oklahoma 73019, USA}

\begin{abstract}
We propose to simulate bosonic pair creation using large arrays of long-lived dipoles with multilevel internal structure coupled to an undriven optical cavity. Entanglement between the atoms, generated by the exchange of virtual photons through a common cavity mode, grows exponentially fast and is described by two-mode squeezing of effective bosonic quadratures. The mapping between an effective bosonic model and the natural spin description of the dipoles allows us to realize the analog of optical homodyne measurements via straightforward global rotations and population measurements of the electronic states, and we propose to exploit this for quantum-enhanced sensing of an optical phase (common and differential between two ensembles). We discuss a specific implementation based on Sr atoms and show that our sensing protocol is robust to sources of decoherence intrinsic to cavity platforms. Our proposal can open unique opportunities for next-generation optical atomic clocks.
\end{abstract}

\maketitle 

The generation of robust and scalable quantum squeezing on an optical transition has the potential to vastly improve optical frequency standards in state-of-the-art atomic clocks and significantly advance capabilities in a variety of fields ranging from gravimetry~\cite{chu2021quantum} to fundamental physics~\cite{ahmed2018quantum,backes2021quantum, arvanitaki2015searching, derevianko2014hunting,kolkowitz2016gravitational,godun2014frequency, sanner2019optical, roberts2020search}. Despite this promise, experiments have not yet used squeezing to make any practical improvements to optical frequency standards in state-of-the-art clocks beyond proof-of-principle experiments~\cite{pedrozo2020entanglement, leroux2010orientation}, due to a variety of physical and technical challenges.
As a result, the development and theoretical study of new proposals to generate scalable, robust entangled states that simultaneously minimize experimental complexity and respect technical constraints are a crucial task to make concrete progress in the field of quantum-enhanced metrology \cite{Hu2019,RLS_TSS_2018, zhang2017cavity, lewis2020protocol, Parkins2017, Davis_2016, davis2017advantages, rls_su2_2021, chu2021quantum}.

In this Letter, we introduce two-mode squeezing (TMS) of atoms in a cavity-QED platform as a feasible path to quantum-enhancement of state-of-the-art clocks. TMS is realized as the result of a process that produces entangled pairs of particles. Prior realizations of TMS in photonic systems generated entangled photons via, e.g., four-wave mixing (FWM) of optical fields in a nonlinear medium \cite{agarwal2012quantum}. Complementary efforts in ultracold bosonic gases have used the intrinsic nonlinearity provided by contact interactions \cite{bucker2011twin,bonneau2013tunable,Zou2018,Hodgman2017,Jaskula2010,Hu2019,Borselli2021}, including spinor Bose-Einstein condensates (BECs) \cite{gross2011atomic,lucke2011twin,bookjans2011,Black2007,zhao2014,qu_2020,Kim2021}, wherein spin-changing collisions between atoms of different internal spin states simulate pair production analogous to degenerate FWM of optical fields~\cite{polzik2016entanglement}. Experiments have also generated TMS via quantum non-demolition (QND) measurements of thermal gases~\cite{vasilakis2015generation, appel2009mesoscopic, schleier2010states, bohnet2014reduced, sewell2012magnetic, bao2020spin}. However, these experiments have all been challenged by a variety of factors -- short interaction times in photonics, finite detection efficiency in QND measurements, and complex spatial dynamics in spinor BECs that limits their scalability~\cite{law1998, jie2020}.

We propose to simulate a pair production process through light-mediated interactions between atoms confined in an undriven optical cavity and exploit it for quantum-enhanced metrology. Pairs of entangled excitations are generated by the exchange of virtual photons between a quartet of internal spin states coupled to a common, far-detuned cavity mode, in a process analogous to FWM. The pinning of the atoms in a deep optical lattice supported by the cavity, in combination with the global range of the effective interaction~\cite{norcia2018cavity}, avoids undesirable motional decoherence and can enable the study of large systems. The exploitation of light-mediated unitary exchange interactions realized by coupling an optical transition to an undriven cavity complements prior work involving Raman transitions \cite{Parkins2017,davis2019photon}, and avoids potential sources of technical noise introduced by driving the cavity with an external field, such as fluctuations in the drive intensity or detuning. Nevertheless, our proposal can in principle be extended to these systems.

TMS states are well known to have excess quantum fluctuations in the phase and amplitude quadratures of each bosonic mode, but suppressed fluctuations of combined quadratures of the two modes. We demonstrate that, despite the complexity of our underlying physical system, the generated quadrature squeezing can be readily accessed and exploited for quantum-enhanced sensing by a sequence of rotations and population measurements that are straightforward to implement and shown to be equivalent to a standard Ramsey sequence. Moreover, engineering TMS on two distinct long-lived optical transitions lets us design protocols to sense 
both differential and sum phases imprinted on the atoms. This is a distinct advantage of our scheme compared to, e.g., atomic homodyne techniques that have been developed in spinor BECs but require simultaneous mixing of multiple internal states \cite{gross2011atomic,peise2015satisfying, hamley2012spin}, or alternatively nonlinear readout protocols to exploit the correlated noise \cite{linnemann2016quantum}. We verify that our protocol is robust to typical sources of decoherence in cavity-QED realizations and thus can be immediately relevant for state-of-the-art time and frequency standards.

\begin{figure}[t]
\centering
\includegraphics[width=1.0\columnwidth]{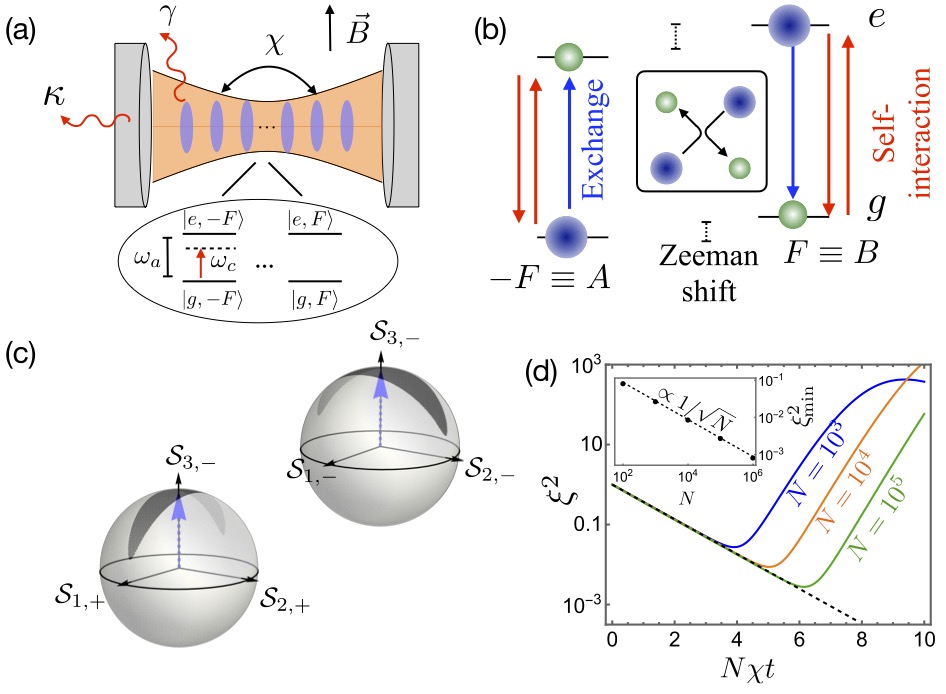}
\caption{(a) Schematic of cavity implementation: interactions ($\chi$) between multilevel atoms (internal structure shown in inset) are mediated by exchange of virtual photons through a common cavity mode of angular frequency $\omega_c = \omega_a + \Delta$ where $\Delta$ is the detuning and $\omega_a$ is the atomic transition angular frequency. The cavity leaks photons through the mirrors at rate $\kappa$ and the atoms undergo spontaneous emission at rate $\gamma$. A magnetic field perpendicular to the cavity axis provides a Zeeman shift, and sets the quantization direction. (b) Possible exchange processes (blue) and self-interactions (red) caused by cavity-mediated interactions. (c) Visualization of the spin squeezing generated during the dynamics, in the combined basis of the $A$ and $B$ manifolds. Blue arrows label the Bloch vector. (d) Squeezing, quantified by the normalized variance $\xi^2 = \frac{4}{N}(\Delta\mathcal{S}_{1,-})^2 = \frac{4}{N}(\Delta\mathcal{S}_{2,+})^2$, for $\delta = N\chi/2$ and different atom numbers $N$. The UPA prediction (dashed line) agrees with TWA calculations (solid lines) until corrections beyond UPA become important (see text). The minimum squeezing is $\xi_{\rm min}^2 \approx 0.88/\sqrt{N}$ as shown in the inset.}
\label{fig1}
\end{figure}

\emph{Engineered FWM. --} We consider an ensemble of atoms trapped by a deep one-dimensional magic optical lattice within an optical cavity, such that the spatial dynamics are effectively frozen. A single cavity mode, with angular frequency $\omega_c$ and power decay linewidth $\kappa$, couples to a long-lived optical transition, with angular frequency $\omega_a$ and natural decay rate $\gamma\ll \kappa$, between a manifold of ground ($g$) and excited ($e$) states with single-photon Rabi frequency $2g_0$ [see Fig.~\ref{fig1}(a)]. We focus on the far-detuned limit, $\vert \Delta \vert = \vert \omega_c - \omega_a \vert \gg g_0\sqrt{N},\kappa$, for $N$ atoms, where the dynamics is near unitary. The cavity field can be adiabatically eliminated and serves only to mediate effective interactions between the atoms \cite{muniz2020exploring}. For concreteness of the following, we consider a system based on the Zeeman levels of the $^1$S$_0$ ($g$) and $^3$P$_0$ ($e$) electronic states in $^{87}$Sr with $F = 9/2$, which are separated by an optical transition frequency forming the basis of state-of-the-art optical lattice clocks \cite{ludlow2015optical}. However, our discussion can be generalized to alternative implementations using, e.g., spatially divided ensembles to emulate the multiple internal transitions \cite{davis2019photon,periwal2021programmable}.

We assume the atomic ensemble is prepared with an equal population of atoms in the electronic states $\ket{g,m=-9/2}$ and $\ket{e,m=9/2}$ where $m$ labels the spin projection of the Zeeman sublevel along the quantization axis set to be perpendicular to the cavity axis (e.g.~by an external magnetic field). This initial state can be prepared using a combination of optical pumping and state-resolved transitions, as explained in ~\cite{SM}. Therein (see also Fig.~\ref{fig1}(b)), we also verify that under this initial condition, the full atomic spin Hamiltonian involving all 20 levels in $^{87}$Sr dominantly drives cavity-mediated dynamics only in the quartet of states, $\ket{g,m=\pm9/2}$ and $\ket{e,m=\pm9/2}$, as the population of other Zeeman sublevels is suppressed by a combination of collective effects and favorable Clebsch-Gordan coefficients \cite{norcia2016superradiance,SM}.The atomic evolution is then described by the effective spin Hamiltonian \cite{SM,muniz2020exploring},
\begin{equation} \label{eqn: two spin ensembles}
\hat H = \hbar\chi \left(\hat S_A^+ + \hat S_B^+\right)\left(\hat S_A^- + \hat S_B^-\right) + \hbar\delta \left(\hat S_B^z - \hat S_A^z\right) .
\end{equation}
Hereafter, we denote the $m=-9/2 (9/2)$ manifold as the $A(B)$ ensemble. We have introduced collective operators $\hat S_{A}^+ = \sum_{i=1}^N \ket{e,A}_i\bra{g,A}_i$, $\hat S_{B}^+ =- \sum_i \ket{e,B}_i\bra{g,B}_i$ and $\hat{S}_{\alpha}^z = 1/2 \sum_i \left( \ket{e,\alpha}_i\bra{e,\alpha}_i - \ket{g,\alpha}_i\bra{g,\alpha}_i\right)$ for $\alpha = A,B$, where the summation runs over all $N$ atoms. The sign convention for the $B$ ensemble accounts for the differing sign of Clebsch-Gordan coefficients for the relevant transitions in each ensemble, which we absorb in the raising and lowering operators rather than Hamiltonian definition for convenience. The cavity detuning $\Delta$ controls the strength of the interaction, $\chi \approx -g_F^2/\Delta$ where the adjusted Rabi frequency $2g_F = 2g_0\sqrt{F/(F+1)}$ includes an additional factor $\sqrt{F/(F+1)}$ arising from Clebsch-Gordan coefficients \cite{SM}. A relative Zeeman shift, $\propto \delta$, splits the energies of the two ensembles. 

The first term of Eq.~\eqref{eqn: two spin ensembles} is a flip-flop process that includes: i) an exchange of an excitation between the $A$ and $B$ ensembles, e.g., $\hat{S}^+_A\hat{S}^-_B + H.c.$, and ii) a self-interaction $\hat{S}^+_A\hat{S}^-_A + \hat{S}^+_B\hat{S}^-_B$. Both can be understood as the simultaneous destruction of a pair of particles in two atomic levels and subsequent creation of a pair in two levels, which is analogous to the process of FWM familiar from quantum and atom optics. We rigorize this analogy by defining Schwinger boson operators $\ha_{g,\alpha}$ and $\ha_{e,\alpha}$ via $\hat S_\alpha^+ = \ha_{e,\alpha}\+ \ha_{g,\alpha}^{\phantom\dagger}$ to rewrite the spin Hamiltonian as,
\begin{align}\label{eqn:FWM}
\hat H_{\rm FWM} &= \hbar\chi \left(\ha_{e,A}\+ \ha_{g,A}^{\phantom\dagger} + \ha_{e,B}\+ \ha_{g,B}^{\phantom\dagger}\right)\left(\ha_{g,A}\+ \ha_{e,A}^{\phantom\dagger} + \ha_{g,B}\+ \ha_{e,B}^{\phantom\dagger}\right)\nonumber\\
&+ \frac{\hbar\delta}{2} \left( \ha_{g,A}\+ \ha_{g,A}^{\phantom\dagger} + \ha_{e,B}\+ \ha_{e,B}^{\phantom\dagger} - \ha_{e,A}\+ \ha_{e,A}^{\phantom\dagger} - \ha_{g,B}\+ \ha_{g,B}^{\phantom\dagger} \right) ,
\end{align}
where the first line describes a set of FWM processes. 
The Hamiltonian (\ref{eqn:FWM}) is closely related to that realized via spin-changing interactions in spin-$1$ BECs \cite{SM} under the assumption that all atoms are restricted to a single common spatial mode. This assumption is not required in cavity-QED implementations \cite{norcia2018cavity,muniz2020exploring,davis2019photon} wherein the infinite-range interactions are generated by a uniform atomic coupling to a single common cavity mode, achieved by selective loading of the atoms in the spatial lattice or by adopting a ring cavity configuration.

\emph{Dynamics of pair creation. --} In quantum optics it is common to make an undepleted pump approximation (UPA) \cite{gerry_knight_2004} to study FWM, corresponding to replacing $\ha_{g,A}, \ha_{e,B} \sim \sqrt{N/2}$ in $\hat H_{\rm FWM}$. For simplicity, we have assumed that the two pump modes are equally populated and treat the general case with unequal pump populations in~\cite{SM}. Further assuming $\delta = N\chi/2$ to effectively remove the mean-field interaction shift due to the self-interaction terms $\chi(\hat{S}^+_A\hat{S}^-_A + \hat{S}^+_B\hat{S}^-_B)$ \cite{SM} we obtain
\begin{align}\label{eqn: TMS}
\hat H_{\rm TMS} &= \frac{N\hbar\chi}{2} \left(\ha_{e,A}\+ \ha_{g,B}\+ + {\rm H.c.}\right). \end{align}
This final form elucidates a resonant production of pairs of bosons, or equivalently the correlated transfer of atom pairs to the internal levels $\vert e,A\rangle$ and $\vert g,B\rangle$. Using $\hat H_{\rm TMS}$, the number of entangled particles is $\nbar(t) = \langle \hat{a}^{\dagger}_{e,A} \hat{a}_{e,A}^{\phantom\dagger} + \hat{a}^{\dagger}_{g,B} \hat{a}_{g,B}^{\phantom\dagger} \rangle = 2\sinh^2 (N\chi t/2)$~\cite{lvovsky2015squeezed, qu_2020}. We benchmark the pair production and other predictions by Eq.~\eqref{eqn: TMS} against the exact dynamics produced by Eq.~\eqref{eqn: two spin ensembles}  in~\cite{SM}.

\emph{Two-mode squeezing for enhanced metrology with an optical transition. --}
It is well established in quantum optics that the Hamiltonian Eq.~\eqref{eqn: TMS} generates squeezing of combined two-mode quadrature fluctuations~\cite{walls_quantum_2008}. Considering $\chi>0$ without loss of generality, within the UPA $\hat H$ produces squeezing along two bosonic quadratures labeled ${Y}_+$ and ${X}_-$ and antisqueezing along conjugate quadratures ${X}_+$ and ${Y}_-$~\cite{SM}, with exponentially fast suppression or growth of the associated quantum noise $ (\Delta X_\pm)^2 = \frac{1}{2}e^{\pm N\chi t}$ and $ (\Delta Y_\pm)^2 = \frac{1}{2}e^{\mp N\chi t}$. In our proposal, the two-mode quadrature squeezing can be observed in collective spin operators that act on our four-level system. Specifically, the squeezed quadratures can be directly mapped to a combination of spin operators, $\sqrt{\frac{N}{2}} \hat X_- = \mathcal{\hat S}_{1,-} \equiv \hat S_B^x - \hat S_A^y$ and $\sqrt{\frac{N}{2}} \hat Y_+ = \mathcal{\hat S}_{2,+} \equiv \hat S_B^y + \hat S_A^x$. Correspondingly, the antisqueezed quadratures are $\sqrt{\frac{N}{2}} \hat Y_- = \mathcal{\hat S}_{2,-} \equiv \hat S_B^y - \hat S_A^x$ and $\sqrt{\frac{N}{2}} \hat X_+ = \mathcal{\hat S}_{1,+} \equiv \hat S_B^x + \hat S_A^y$.

We can visualize the squeezed quantum noise of the combined spin state corresponding to the $A$ and $B$ transitions on a pair of coupled Bloch spheres defined by axes $(\mathcal{S}_{1,-},\mathcal{S}_{2,-},\mathcal{S}_{3,-})$ and $(\mathcal{S}_{1,+},\mathcal{S}_{2,+},\mathcal{S}_{3,-})$ that share a common vertical component $\mathcal{S}_{3,-} = S_B^z - S_A^z$ and for which the corresponding operators obey standard SU(2) commutation relations \cite{SM}. As shown in Fig.~\ref{fig1}(c), the state is squeezed in both Bloch spheres, $(\Delta \mathcal{S}_{1,-})^2 = (\Delta \mathcal{S}_{2,+})^2 = Ne^{-N\chi t}/4$, relative to the level of the initial separable state.

The UPA prediction for the squeezing, $(\Delta\mathcal{S}_{1,-})^2$ and $(\Delta\mathcal{S}_{2,+})^2$, is verified in Fig.~\ref{fig1}(d) by comparing to a calculation of the variances based on a numerical simulation of the full multilevel cavity implementation, and we find excellent agreement up to $\nbar \sim 0.76 \sqrt{N}$. The multilevel cavity dynamics are obtained using a truncated Wigner approximation (TWA), which approximates the quantum dynamics by averaging over an ensemble of classical trajectories with initial conditions chosen to reproduce the quantum fluctuations of the initial state \cite{Zhu_2019,Lepoutre,Schachenmayer,Patscheider, sundar2019analysis}. We include all possible exchange processes between the complete set of $4F+2$ ground and excited atomic levels in our TWA simulation, including, e.g., those mediated by photons with polarization perpendicular to the quantization axis ~\cite{SM, PineiroOrioli_2022}.

A Ramsey protocol~\cite{ma2011quantum} that uses only collective rotations and population measurements of the spins encoded in the $A$ and $B$ manifolds can be used to take advantage of the squeezing in the Bloch sphere for enhanced sensing of phase shifts imprinted on the optical transition. The protocol is analogous to optical homodyne techniques in quantum optics, as well as atomic homodyne \cite{Oberthaler2014,peise2015satisfying} or measurements of squeezing in spin-$1$ BECs \cite{hamley2012spin,Anders2018}. However, as our atomic realization is based on four internal levels, as opposed to three in spin-$1$ BECs, we do not require any coherent mixing of the $F = \pm 9/2$ manifolds. This also distinguishes our approach from prior demonstrations of interferometry with Dicke-like states realized in spin-$1$ BECs \cite{lucke2011twin}, which treat the $m_F=\pm 1$ modes as the two internal levels of a collective spin-1/2 system and uses a Holland-Burnett-type protocol \cite{HollandBurnett1993}. Such an approach is sensitive to decoherence \cite{Huelga1997} and readout errors \cite{rls_su2_2021}, while in our system it would also add the complex requirement of engineering a coupling between the $F=\pm 9/2$ states.

\begin{figure}[t]
\centering
\includegraphics[width=1.0\columnwidth]{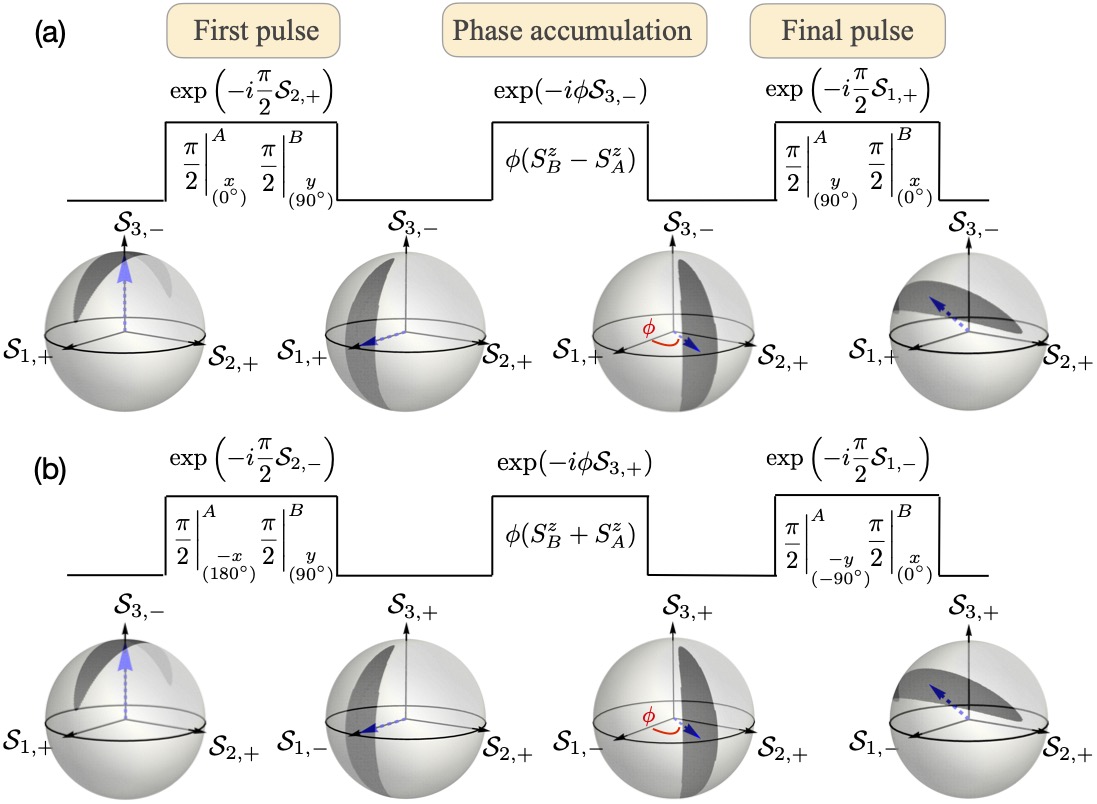}
\caption{(a) Ramsey sequence to measure a differential clock phase shift imprinted on the atoms. A first $\pi/2$ pulse rotates the Bloch vector to the equator. Next, a phase accumulation rotates the Bloch vector about $S^z_B - S^z_A$ by an angle $\phi$, before a final $\pi/2$ pulse rotates the Bloch vector for readout via measuring $\hat S^z_B - \hat S^z_A$. (b) Ramsey sequence to measure the sum phase imprinted on the atoms. The first $\pi/2$ pulse defines the squeezed distribution on a joint Bloch sphere of $A$ and $B$ defined by the axes $(\mathcal{S}_{1,-},\mathcal{S}_{2,+},\mathcal{S}_{3,+})$. After phase accumulation about $S^z_B + S^z_A$, the final pulse again rotates the Bloch vector for readout via measuring $\hat S^z_B + \hat S^z_A$. In both cases, dashed blue arrows mark the Bloch vector at each stage. The subscript $x (y)$ for the pulses denote the axis of rotation $S_x (S_y)$, and the degree $0^\circ (90^\circ)$ in the subscript conveys the same information via the phase of the laser pulse.}
\label{fig2}
\end{figure}

Here, we present Ramsey protocols for measuring sum and difference optical phases imprinted on the atoms, illustrated in Fig.~\ref{fig2}. Measuring differential phases has several applications including gravimetry~\cite{tino2021testing}, measuring gravitational redshifts~\cite{bothwell2021resolving, chou2010optical, grotti2018geodesy, takamoto2020test}, and detecting gravitational waves~\cite{tse2019quantum, acernese2019increasing} and dark matter~\cite{kennedy2020precision}. Measuring the sum phase is useful for improving state-of-the-art optical atomic clocks.

To measure a differential phase imprinted by a rotation about $ S^z_B - S^z_A$, we begin our protocol with a $\pi/2$ pulse that rotates atoms in the $A$ ensemble about $S^x$ and those in the $B$ ensemble by $S^y$, i.e., implements $\exp(-i\frac{\pi}{2} \mathcal{\hat S}_{2,+})$. Next, we accumulate a relative phase shift by a rotation of $\phi$ about $-S^z_A$ and $S^z_B$ (i.e., $\mathcal{S}_{3,-}$), and finally apply a second $\pi/2$ pulse which rotates atoms in the $A$ and $B$ ensembles about $S^y$ and $S^x$, i.e., implements $\exp(-i \frac{\pi}{2}\mathcal{\hat S}_{1,+})$. The action of this pulse sequence on the state is best visualized by looking at the spin distribution on the Bloch spheres, as shown in Fig.~\ref{fig2}(a), where the initial spin distribution is to that of the lower Bloch sphere in Fig.~\ref{fig1}(c). The final pulse converts the rotation $\hat U_\phi$ into a change in the difference in atomic inversions
\begin{equation}
\braket{ \hat S^z_B - \hat S^z_A} = \frac{N}{2}\sin\phi .
\end{equation}
This Ramsey sequence does not imprint any information about the differential phase on the upper Bloch sphere in Fig.~\ref{fig1}(c) \cite{SM}.

The sum phase, imprinted by a collective rotation around $\mathcal{S}_{3,+} \equiv S^z_A + S^z_B$, is similarly inferred by another Ramsey protocol shown in Fig.~\ref{fig2}(b). Note that neither Bloch sphere in Fig.~\ref{fig1}(c) has $\mathcal{S}_{3,+}$ as an axis. Therefore, the first pulse in this Ramsey protocol, implementing $\exp(-i\frac{\pi}{2}\hat{\mathcal{S}}_{2,-})$, is chosen such that it rotates the axes in the Bloch sphere from $(\mathcal{S}_{1,+}, \mathcal{S}_{2,+}, \mathcal{S}_{3,-})$ to $(-\mathcal{S}_{3,+}, \mathcal{S}_{2,+}, \mathcal{S}_{1,-})$, thus introducing $\mathcal{S}_{3,+}$ into relevance. The remainder of the sequence proceeds analogously to that for the differential phase. 

The sensitivity of both protocols, computed within the UPA, is
\begin{equation}\label{eqn: sens UPA}
(\Delta\phi)^2 \equiv \frac{(\Delta O)^2}{ (d\braket{\hat O}/d\phi)^2} = \frac{e^{-N\chi t}}{N} + \frac{\bar{n}(\bar{n}+2)}{4N^2}\tan^2\phi, 
\end{equation}
where $\hat O$ is the observable measured. Equation \eqref{eqn: sens UPA} predicts an advantage relative to the standard quantum limit (SQL), $(\Delta\phi)^2 = 1/N$, which sets the optimal resolution achievable with uncorrelated particles, for any $\bar{n} > 0$ and a wide dynamic range of $\phi$, $|\tan\phi| < 2\sqrt{N}/\nbar$, which can be $\mathcal{O}(1)$.

The sensitivity [Eq.~\eqref{eqn: sens UPA}] can be degraded by other effects not included in the ideal analysis, including (i) corrections beyond the UPA, (ii) decoherence, and (iii) fluctuations  in the total and relative populations of the $A$ and $B$ ensembles. We discuss (i) and (ii) below, and leave the discussion on (iii) to \cite{SM}. There, we show that number fluctuations at the level of shot noise provide a comparable degradation of the sensitivity to that generated by (i) and (ii).

The leading corrections to $\hat H_{\rm TMS}$ can be captured by iteratively modifying the UPA to include depletion of the pump states $\ket{g,A}$ and $\ket{e,B}$ by $\nbar/2$. This is achieved by setting $\ha_{g,A},\ha_{e,B} \approx \sqrt{\frac{N-\nbar}{2}}$ where $\nbar = 2\sinh^2\frac{N\chi t}{2}$ is taken to be the original UPA result as a first approximation. Making this correction has two physical consequences \cite{SM}: (i) the effective nonlinearity $\chi (N-\nbar)/2$ driving pair production is reduced relative to the UPA, and (ii) the pair production is no longer resonant as the Zeeman shift $\delta = \chi N/2$ is static and does not completely cancel the mean-field shift introduced by the self-interaction terms in Eq.~(\ref{eqn:FWM}). For $1 \ll \nbar \ll N$ we then obtain the beyond-UPA sensitivity \cite{SM}
\begin{equation}\label{eqn: sensitivity beyond UPA}
(\Delta\phi)^2 \approx \frac{1}{2N\nbar} + \frac{\nbar^3}{2N^3} + \frac{\nbar(\nbar+2)}{4N^2}\tan^2\phi . 
\end{equation}
The optimal sensitivity remains enhanced relative to the SQL, with a lower bound of $(\Delta\phi)^2 = 2/(3^{3/4}N^{3/2})$ that occurs for $\nbar = \sqrt{N}/3^{1/4}$, in agreement with TWA results.

\begin{figure}[t]
\centering
\includegraphics[width=1.0\columnwidth]{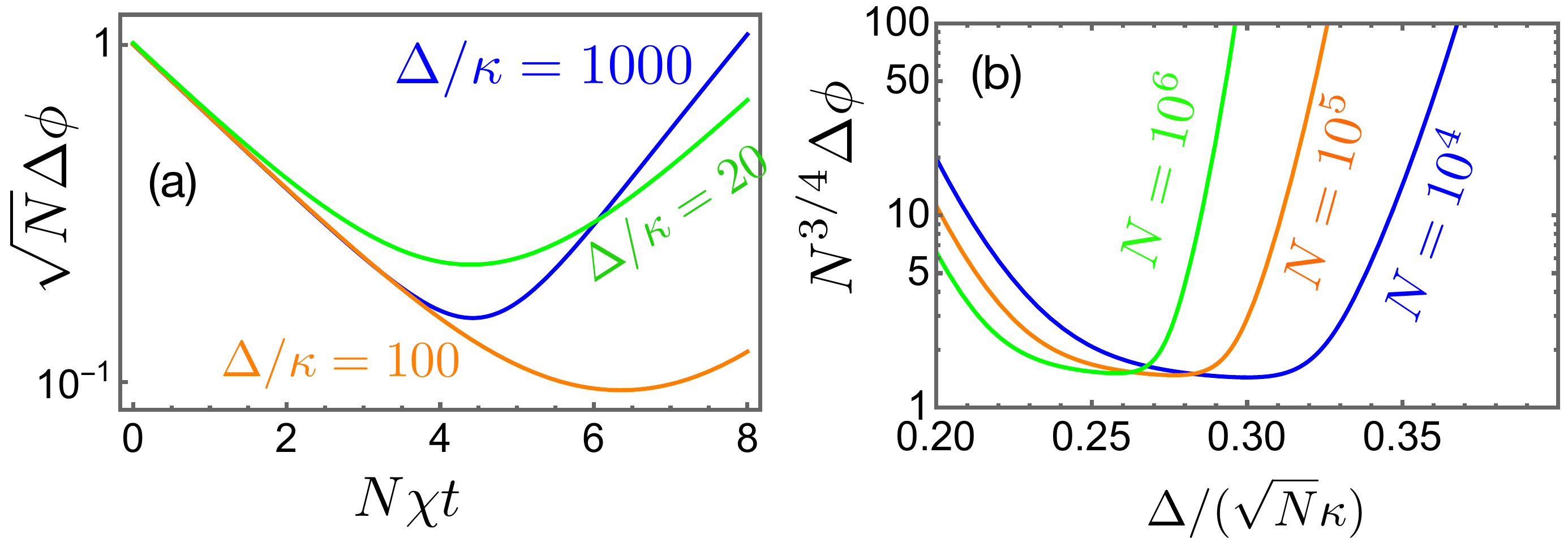}
\caption{(a) Scaled sensitivity $\sqrt{N}\Delta\phi$ versus time for different cavity detunings $\Delta$, and $C=10$ and $N=10^5$. The best sensitivity is achieved at an optimum time for each $\Delta$. (b) Scaled best sensitivity $N^{3/4}\Delta\phi$ versus $\Delta/(\sqrt{N}\kappa)$. In each case, the minima occur at similar $\Delta/(\sqrt{N}\kappa)$, and have similar values (up to logarithmic corrections).}
\label{fig3}
\end{figure}

\emph{Decoherence. --} Dissipative noise in our system intrinsically arises from superradiant decay, at a rate $\Gamma \approx g_F^2\kappa/\Delta^2$, due to leakage of the photons that mediate the effective atom-atom interaction from the cavity, and single particle spontaneous emission into free space at the rate $\gamma$. While both are deleterious for sensing, we show that our protocol can achieve sub-SQL sensitivity even with these sources of decoherence.

Collective decay is treated by solving a Lindblad master equation with jump operator $\hat L = \sqrt{\Gamma} (\hat S_A^- + \hat S_B^-)$, which captures the dominant process where an emitted photon polarized along the quantization axis is lost from the cavity \cite{SM}. Spontaneous emission is included through jump operators $\hat L_i = \sqrt{\gamma}(\hat \sigma_{A,i}^- + \hat \sigma_{B,i}^-)$, where $\sigma_{\alpha,i}^-$ is the spin-lowering operator in the $\alpha = $A,B manifold for the $i$th atom.

In Fig.~\ref{fig3}(a) we plot the scaled sensitivity $\sqrt{N}\Delta\phi$ as a function of time in the presence of decoherence for a range of cavity detunings. It decreases from the initial SQL to a minimum value at an optimum time that depends on the detuning. 
This optimum time balances the gain obtained by reaching a higher $\nbar$ versus the loss in squeezing due to decoherence. Optimizing this interplay via $\Delta$ (thus tuning $\chi$ relative to $\Gamma$ and $\gamma$), we obtain a best achievable sensitivity [see Fig.~\ref{fig3}(b) and \cite{SM}]
\begin{align}\label{eqn: best sensitivity}
(\Delta\phi)^2 = &\frac{\sqrt{2\ln(2NC)}}{N^{3/2}\sqrt{C}},
\end{align}
for $\Delta = \frac{\kappa\sqrt{NC}}{2\sqrt{\ln(2NC)}}$, with $C = 4g_F^2/\kappa\gamma$ the single-atom cooperativity. Current experimental setups~\cite{norcia2016superradiance, norcia2018cavity}, with $\kappa/2\pi \sim 150$ kHz and $g/2\pi \sim 4$ Hz, can reach a collective cooperativity $NC=4\times 10^5$ with $N=10^6$ atoms, and $NC=10^7$ is within reach. The sensitivity in Eq.~\eqref{eqn: best sensitivity} is only slightly degraded from the ideal case [see Fig.~\ref{fig1}(d) and Eq.~\eqref{eqn: sensitivity beyond UPA}] and is competitive with the best sensitivities achievable with the paradigmatic approach of one-axis twisting when decoherence is incorporated \cite{Hu2017,RLS_TSS_2018,chu2021quantum}.

\emph{Outlook. --} Our proposal offers new opportunities to study and exploit the exponentially rapid generation of entanglement in atomic systems, driven by connections to well established concepts in quantum optics. Moreover, our proposal highlights new possibilities for the realization and investigation of diverse models of bosonic pair production in highly tunable quantum simulators featuring spin-spin interactions.

\begin{acknowledgments}
We thank M. Affolter, A. Carter, and T. Bilitewski for a careful reading and comments on the manuscript. AMR acknowledges support from the Air Force Office of Scientific Research Grants No. FA9550-18-1-0319 and No. FA9550-19-1-0275, from the Defense Advanced Research Projects Agency and Army Research Office Grant No. W911NF-16-1-0576, by the National Science Foundation JILA-Physics Frontier Center Grant No. PHY-1734006, Quantum Leap Challenge Institute - Office of Multidisciplinary Activities Grant No. 2016244, by the United States Department of Energy, Office of Science, National Quantum Information Science Research Centers Quantum Systems Accelerator, and from the National Institute of Standards and Technology. RLS acknowledges support from the National Science Foundation Grant No. PHY-2110052.
\end{acknowledgments}

\bibliography{refs}

\end{document}


\renewcommand{\theequation}{S.\arabic{equation}}
\renewcommand{\thefigure}{S.\arabic{figure}}
\renewcommand{\thesection}{S.\arabic{section}}

\title{Supplementary Material: Bosonic pair production and squeezing for optical phase measurements in long-lived dipoles coupled to a cavity}

\author{Bhuvanesh Sundar}
\affiliation{Center for Theory of Quantum Matter, University of Colorado, Boulder, CO 80309, USA}
\affiliation{JILA, NIST, Department of Physics, University of Colorado, Boulder, CO 80309, USA}
\author{Diego Barberena}
\affiliation{Center for Theory of Quantum Matter, University of Colorado, Boulder, CO 80309, USA}
\affiliation{JILA, NIST, Department of Physics, University of Colorado, Boulder, CO 80309, USA}
\author{Asier Pi\~neiro Orioli}
\affiliation{Center for Theory of Quantum Matter, University of Colorado, Boulder, CO 80309, USA}
\affiliation{JILA, NIST, Department of Physics, University of Colorado, Boulder, CO 80309, USA}
\author{Anjun Chu}
\affiliation{Center for Theory of Quantum Matter, University of Colorado, Boulder, CO 80309, USA}
\affiliation{JILA, NIST, Department of Physics, University of Colorado, Boulder, CO 80309, USA}
\author{James K. Thompson}
\affiliation{JILA, NIST, Department of Physics, University of Colorado, Boulder, CO 80309, USA}
\author{Ana Maria Rey}
\affiliation{Center for Theory of Quantum Matter, University of Colorado, Boulder, CO 80309, USA}
\affiliation{JILA, NIST, Department of Physics, University of Colorado, Boulder, CO 80309, USA}
\author{Robert J. Lewis-Swan}
\affiliation{Homer L. Dodge Department of Physics and Astronomy, The University of Oklahoma, Norman, OK 73019, USA}
\affiliation{Center for Quantum Research and Technology, The University of Oklahoma, Norman, OK 73019, USA}

\maketitle

\section{Proposed experimental implementation}\label{sec: implementation}
In the main text we introduced a proposal to realize an analog of bosonic four-wave mixing by coupling multilevel atoms to a far-detuned optical cavity mode. Here, we present a detailed discussion of the full atom-light Hamiltonian describing this system and show that it can be reduced to an effective spin model, which is presented as Eq.~(1) in the main text.

\subsection{Atom-light model}
We consider multilevel atoms with a ground manifold and an excited manifold separated by an energy $\hbar\omega_a$, and $2F+1$ Zeeman levels in each manifold. We denote the excited states as $\ket{e,m}$ and the ground states as $\ket{g,m}$ where the index $m \in [-F,F]$ specifies the Zeeman level. The cavity supports a pair of photon modes with degenerate angular frequency $\omega_c$ but different polarization. Without loss of generality, we take one of the cavity modes to be linearly polarized along the atomic quantization axis, which we call $\Pi$-polarization, and the other cavity mode to be linearly polarized perpendicular to the quantization axis, which we call $\Sigma$-polarization. The atoms couple to these two cavity modes with single-photon Rabi frequency $2g_0$.

The dynamics of the atom-light system is modeled by the Lindblad master equation,
\begin{equation}\label{eqn: master}
\hbar\frac{d\rho}{dt} = -i[\hat H_{\rm tot}, \rho] + \mathcal{L}_c[\rho] + \mathcal{L}_s[\rho].
\end{equation}
Here, $\hat H_{\rm tot} = \hat H_A + \hat H_L + \hat H_{AL}$ is a Hamiltonian including contributions from the atoms, cavity modes, and atom-light coupling:
\begin{align}
& \hat H_A = \hbar\omega_a \hat n_e + \hbar(\delta_g \hat F^z_g + \delta_e \hat F^z_e),\nonumber\\
& \hat H_L = \hbar\omega_c \sum_{\alpha=\Pi,\Sigma} \hc_\alpha\+ \hc_\alpha^{\phantom\dagger},\nonumber\\
& \hat H_{AL} = \hbar g_0 \left( \hc_\Pi \hat\Pi^+ + \hc_\Sigma \hat\Sigma^+ + {\rm h.c.} \right).
\end{align}
In the above equations, $\hat n_e$ is the occupation in the excited manifold, $\hat F^z_{g(e)} = \sum_{i=1}^N \sum_{m=-F}^F m\ket{g(e),m}_i\bra{g(e),m}$ is the azimuthal spin operator for the ground (excited) manifold, and $\hc_\alpha$ annihilates a photon in the cavity mode with polarization $\alpha$. The terms $\hbar\delta_g \hat F^z_g$ and $\hbar\delta_e \hat F^z_e$ arise from Zeeman shifts due to an applied magnetic field. The operators $\hat\Sigma^+ (\hat\Sigma^-)$ and $\hat\Pi^+ (\hat\Pi^-)$ are collective atomic operators that excite (de-excite) atoms by absorbing a $\Sigma$-polarized and $\Pi$-polarized cavity photon, respectively, given by $\hat\Pi^+ = \sum_{i,m} C_m^0\ket{e,m}_i\bra{g,m}$ and $\hat\Sigma^+ = i \sum_{i,m,q=\pm1} C_m^q \ket{e,m+q}_i\bra{g,m}/\sqrt{2}$, where $C_m^q = \braket{F,m; 1,q \vert F,m+q}$ is the Clebsch-Gordan coefficient associated with exciting from $\ket{g,m}$ to $\ket{e,m+q}$. 

It is convenient to move to a rotating frame that rotates at the atomic frequency $\omega_a$. In this frame, the atomic angular frequency and cavity frequency are shifted by $\omega_a$, yielding the Hamiltonian
\begin{align}
\hat H_{\rm tot} = & \hbar(\delta_g \hat F^z_g + \delta_e \hat F^z_e) + \hbar\Delta\sum_{\alpha=\Pi,\Sigma} \hc_\alpha\+ \hc_\alpha^{\phantom\dagger} \nonumber\\
& + \hbar g_0 \left( \hc_\Pi \hat\Pi^+ + \hc_\Sigma \hat\Sigma^+ + {\rm h.c.} \right),
\end{align}
and we define the detuning of the cavity from the atomic transition, $\Delta = \omega_c - \omega_a$.

There are two Lindblad terms in the master equation \eqref{eqn: master}, $\mathcal{L}_c[\rho]$ and $\mathcal{L}_s[\rho]$ that describe decoherence of the photon and atomic degrees of freedom. The former term captures leakage of photons out of the cavity at rate $\kappa$, and is given by
\begin{equation}
\mathcal{L}_c[\rho] = \hbar\kappa\sum_{\alpha=\Pi,\Sigma} \left(\hc_\alpha^{\phantom\dagger} \rho \hc_\alpha\+ - \frac{1}{2}\hc_\alpha\+\hc_\alpha^{\phantom\dagger} \rho - \frac{1}{2}\rho\hc_\alpha\+\hc_\alpha^{\phantom\dagger}\right) .
\end{equation}
The second Lindblad term, $\mathcal{L}_s[\rho]$, models spontaneous decay of atoms from the excited manifold at rate $\gamma$,
\begin{equation}
\mathcal{L}_s[\rho] = \sum_{i=1}^N \sum_{l=\pi,\sigma_\pm} \left(\hat L_{l,i}^- \rho \hat L_{l,i}^+ - \frac{1}{2}\hat L_{l,i}^+\hat L_{l,i}^- \rho - \frac{1}{2}\rho\hat L_{l,i}^+\hat L_{l,i}^-\right).
\end{equation}
This expression is further decomposed into three kinds of Lindblad jump operators: $\hat L_{\pi,i}^- = \sqrt{\hbar\gamma} \sum_m C_m^0 \ket{g,m}_i \bra{e,m}$ for spontaneous decay that preserves magnetization, and $\hat L_{\sigma_\pm,i}^- = \sqrt{\hbar\gamma} \sum_m C_m^{\pm 1} \ket{g,m}_i \bra{e,m \pm 1}$ for spontaneous decay that leads to a change in magnetization by $\pm 1$.

\subsection{Effective spin model}\label{sec:eff_spin_model}
When the cavity is detuned sufficiently far from the atomic transition, $|\Delta| \gg g_0 \sqrt{N}$, we can adiabatically eliminate the cavity photons and obtain an effective master equation for the atoms, $\hbar\frac{d\rho}{dt} = -i[\hat H, \rho] + \mathcal{L}_c[\rho] + \mathcal{L}_s[\rho]$ \cite{muniz2020exploring}. The effective Hamiltonian is
\begin{equation} \label{eqn: H}
\hat H = \hbar\chi_0 \left( \hat\Sigma^+ \hat\Sigma^- + \hat\Pi^+ \hat\Pi^- \right) + \hbar(\delta_g \hat F^z_g + \delta_e \hat F^z_e),
\end{equation}
where $\chi_0 \simeq -g_0^2/\Delta$ and the Lindblad terms are
\begin{align}\label{eqn: Lindblads}
\mathcal{L}_c[\rho] = &\hbar\Gamma_0 \left(\hat\Pi^+ \rho \hat\Pi^- - \frac{1}{2}\hat\Pi^+\hat\Pi^- \rho - \frac{1}{2}\rho\hat\Pi^+\hat\Pi^-\right) \nonumber\\
&+ \hbar\Gamma_0 \left(\hat\Sigma^+ \rho \hat\Sigma^- - \frac{1}{2}\hat\Sigma^+\hat\Sigma^- \rho - \frac{1}{2}\rho\hat\Sigma^+\hat\Sigma^-\right), \nonumber\\
\mathcal{L}_s[\rho] = &\sum_{i=1}^N \sum_{l=\pi,\sigma_\pm} \left(\hat L_{l,i}^- \rho \hat L_{l,i}^+ - \frac{1}{2}\hat L_{l,i}^+\hat L_{l,i}^- \rho - \frac{1}{2}\rho\hat L_{l,i}^+\hat L_{l,i}^-\right),
\end{align}
where $\Gamma_0 \approx g_0^2\kappa/\Delta^2$. The terms in $\hat H$ proportional to $\chi_0$ are cavity photon-mediated exchange of atomic excitations, and the terms in $\mathcal{L}_c[\rho]$ capture collective decay of atoms from the excited manifold, also called superradiant decay.

As discussed in the main text, we consider initial conditions where half of the atoms are in $\ket{g,m=-F}$ and half in $\ket{e,m=F}$. Thus of all the exchange processes mediated by the cavity photons and written in Eq.~\eqref{eqn: H}, the dominant exchange occurs in the $m=-F$ and $m=F$ sublevels, mediated by $\Pi$-polarized photons. The reason for this is that the Clebsch-Gordan coefficients are largest for these processes, $C_{\pm F}^0 = \pm \sqrt{F/(F+1)}$. The exchange process with the next highest amplitude, that exchanges atoms from $\ket{g,-F}\ket{e,F}$ to $\ket{e,-F+1}\ket{g,F-1}$ mediated by the exchange of $\Sigma$-polarized photon, has a smaller amplitude due to a smaller Clebsch-Gordan coefficient, $(C_F^{-1})^2 = 1/(F+1)$. Therefore, we expect that the populations in $\ket{e,-F+1}$ and $\ket{g,F-1}$ grow at an exponentially slower rate than in $\ket{e,-F}$ and $\ket{g,F}$. The amplitude of the process mediated by the $\Pi$-polarized photons, which exchanges atoms from $\ket{g,-F}\ket{e,F}$ to $\ket{e,-F}\ket{g,F}$, is proportional to $N\chi_0 (C_F^0)^2$. At early times, the amplitudes of all the other exchange processes are at most $O(\sqrt{N}\chi_0)$.

Thus, keeping only the dominant exchange terms, from Eq.~(\ref{eqn: H}) we obtain the effective spin model [Eq.~(1)] of the main text. The spin raising and lowering operators in Eq.~(1) in the main text are the raising and lowering operators in $\Pi^\pm$ projected into the $A$ and $B$ manifolds, i.e. $\hat S_A^+ = \sum_i \ket{e,-F}_i\bra{g,-F}$, $\hat S_B^+ = -\sum_i \ket{e,F}_i\bra{g,F}$, and the inversion operators are $\hat S_A^z = \sum_i (\ket{e,-F}_i\bra{e,-F} - \ket{g,-F}\ket{g,-F})/2$ and $\hat S_B^z = \sum_i (\ket{e,F}_i\bra{e,F} - \ket{g,F}\ket{g,F})/2$. The Clebsch-Gordan coefficients that appeared in $\hat\Pi^\pm$ now do not appear in the spin operators $\hat S_{A,B}^\pm$, but are instead included the definition of the parameters in Eq.~(1) as $\chi \approx -g_F^2/\Delta$, where $g_F = g_0\sqrt{F/(F+1)}$, and $\delta = F(\delta_e - \delta_g)$. 

To justify our argument that the effective spin model [Eq.(1) in the main text] is the full spin Hamiltonian [Eq.~\eqref{eqn: H}] projected into the $m=\pm F$ manifolds, we demonstrate that a numerical simulation of the dynamics of the full spin model [Eq.~\eqref{eqn: H}] with all $2F+2$ spin levels produces results consistent with an analytical prediction of the dynamics for the reduced effective spin model. In the next section, we describe how to analytically model the dynamics with an equivalent bosonic description of Eq.(1) and a subsequent approximation. In the section following that, we describe the numerical method that we use to simulate the full spin model [Eq.(1)].

\section{Preparing the atoms in two ensembles}
Our protocol requires us to prepare $N/2$ atoms each in $\ket{g,-F}$ and $\ket{e,F}$. Previous experiments with $^{87}$Sr have prepared atoms in $\ket{g,m=\pm F}$ using optical pumping~\cite{norcia2018cavity, norcia2018frequency}. 
This was achieved via  narrow-line laser cooling on the $^1S_0 \rightarrow ^3\hspace{-0.12cm}P_1$ transition, which is a common choice. If $\pi$-polarized light is used to interrogate the $^1S_0(F = 9/2) \rightarrow ^3\hspace{-0.12cm}P_1(F' = 7/2)$ transition, the $m_F = \pm 9/2$ states in the ground manifold are the dark states, and atoms will accumulate there at the end of cooling process. Due to symmetric Clebsch-Gordan coefficients, the populations in the ground $\ket{m_F=\pm 9/2}$ states are roughly $N/2$ each. This state preparation has bee previously demonstrated in~\cite{norcia2018frequency}. A final $\pi$ pulse acting only on the $m_F=+9/2$ $^1S_0 \rightarrow ^3\hspace{-0.12cm}P_0$ transition, i.e. $\ket{g,F} \rightarrow \ket{e,F}$, using the differential g-factor between the $^1S_0 \rightarrow ^3\hspace{-0.12cm}P_0$ levels  as demonstrated in Ref.~\cite{boyd2006optical}, prepares the required initial state. This preparation protocol  may lead to  a statistical atom number imbalance between $\ket{g,-F}$ and $\ket{e,F}$ of the order of $\delta N \sim \sqrt{N}$. We later investigate the effects of this imbalance in more detail.

\section{Dynamics in equivalent bosonic description}\label{sec: unitary UPA}

In the main text, we reported analytic expressions for the internal state occupations $n_1(t)$ and $n_2(t)$ based on the reduction of the multilevel system discussed in Sec.~\ref{sec:eff_spin_model} [and Eq.~(1) of the main text] to an ideal bosonic two-mode squeezing process. Here, we derive these analytic expressions within UPA. The expressions in the main text are for $\delta = N\chi/2$.

We start from the equivalent representation of Eq.~(1) in the main text in terms of Schwinger bosons,
\begin{align}\label{eqn:Hbosonfull}
H = &\hbar\chi(\ha_{e,A}\+ \ha_{g,A}^{\phantom\dagger} + \ha_{e,B}\+ \ha_{g,B}^{\phantom\dagger})(\ha_{g,A}\+ \ha_{e,A}^{\phantom\dagger} + \ha_{g,B}\+ \ha_{e,B}^{\phantom\dagger}) \nonumber\\
& + \frac{\hbar\delta}{2} (\ha_{e,B}\+ \ha_{e,B}^{\phantom\dagger} - \ha_{g,B}\+ \ha_{g,B}^{\phantom\dagger} - \ha_{e,A}\+ \ha_{e,A}^{\phantom\dagger} + \ha_{g,A}\+ \ha_{g,A}^{\phantom\dagger}).
\end{align}
Under the UPA, we replace $\ha_{g,A}\+ \ha_{g,A}^{\phantom\dagger} = \frac{N}{2} - \ha_{e,A}\+ \ha_{e,A}^{\phantom\dagger}$, $\ha_{e,B}\+ \ha_{e,B}^{\phantom\dagger} = \frac{N}{2} - \ha_{g,B}\+ \ha_{g,B}^{\phantom\dagger}$, and $\ha_{g,A}\+ \approx \ha_{e,B}\+ \approx \sqrt{\frac{N}{2}}$. Then, the Hamiltonian (\ref{eqn:Hbosonfull}) reduces to
\begin{align} \label{eqn: upa}
H \approx &\frac{N\hbar\chi}{2}(\ha_{e,A}\+  +  \ha_{g,B}^{\phantom\dagger})( \ha_{e,A}^{\phantom\dagger} + \ha_{g,B}\+ ) \nonumber\\
 & - N\hbar\delta (\ha_{e,A}\+ \ha_{e,A}^{\phantom\dagger} + \ha_{g,B}\+ \ha_{g,B}^{\phantom\dagger}).
\end{align}

Since the Hamiltonian \eqref{eqn: upa} is quadratic in Schwinger boson operators, the Heisenberg time evolution of these operators can be analytically solved by integrating the matrix equation
\begin{equation} \label{eqn: heisenberg eqn unitary}
i\partial_t \left( \begin{array}{c} \hat a_{e,A} \\ \hat a_{g,B}\+ \end{array} \right) = 
\left( \begin{array}{cc} \frac{N\chi}{2} - \delta & \frac{N\chi}{2} \\ -\frac{N\chi}{2} & -\frac{N\chi}{2} + \delta \end{array} \right)
\left( \begin{array}{c} \hat a_{e,A} \\ \hat a_{g,B}\+ \end{array} \right).
\end{equation}
The occupations are then obtained as,
\begin{equation}\label{eqn: occs}
n_{e,A}(t) = n_{g,B}(t) = \frac{ (N\chi)^2 }{4\delta(N\chi-\delta)} \sinh^2 \left( t\sqrt{\delta(N\chi-\delta)} \right) .
\end{equation}
When $\delta = N\chi/2$ the Zeeman shift cancels the mean-field energy shift [corresponding to the elastic interaction terms in \eqref{eqn:Hbosonfull}], making the pair production resonant. Then, the occupations grow the fastest as $n_{e,A}(t) = n_{g,B}(t) = \sinh^2\frac{N\chi t}{2}$.

\section{Numerical benchmarks}

\subsection{Validity of the 4-level model approximation}

In the main text, we approximated the full spin model [Eq.~\eqref{eqn: H}] with a four-level spin model [Eq.~(1)], based on the argument that the smaller amplitudes for scattering particles into Zeeman levels with $m\neq \pm F$ lead to negligible occupations in those levels (see also earlier discussion of the effective spin model). In this section, we verify the validity of this approximation, by comparing the dynamics produced by the full spin model involving many levels [Eq.~\eqref{eqn: H}], with the dynamics produced by the reduced 4-level spin model [Eq.~(1)]. The 4-level spin model is amenable to numerical exact diagonalization (ED) for moderate system sizes [$N \lesssim 10^5$] in the collective manifold of the two ensembles. The full spin model, which involves, e.g., 20 levels for $^{87}$Sr, is not amenable to numerical exact diagonalization. Therefore we calculate the dynamics in that case using a semiclassical method based on the truncated Wigner approximation (TWA).

To be concrete, in the TWA we obtain expectation values of observables by averaging over an ensemble of trajectories, and each trajectory is obtained by integrating classical equations of motion for initial values of operators sampled from a probability distribution which properly accounts for quantum fluctuations~\cite{Polkovnikov_AP2010}.

Specifically, we integrate classical equations of motion for collective multilevel spin operators $\hat S_{\alpha\beta} = \sum_i \ket{\alpha}_i\bra{\beta}_i$, where $\alpha$ and $\beta$ can be a ground state $(g,m)$ or an excited state $(e,m)$. There are $(4F+2)^2$ such spin operators. Their initial values are sampled from a multivariate Gaussian distribution with mean and covariance matrix given by the mean and covariance in the initial state,
\begin{align}
& \mu_{\alpha\beta} = \braket{ \hat S_{\alpha\beta} }, \nonumber\\
& \sigma_{\alpha\beta,\gamma\delta} = \sqrt{ \frac{1}{2} \braket{ \hat S_{\alpha\beta} \hat S_{\gamma\delta} + \hat S_{\gamma\delta} \hat S_{\alpha\beta}  } - \braket{ \hat S_{\alpha\beta} } \braket{ \hat S_{\gamma\delta} } }.
\end{align}
After creating an ensemble of initial values $\{ S_{\alpha\beta}(t=0) \}$, each initial value is then propagated according to their equations of motion (see below).

To obtain the equations of motion, it is convenient to first write the Hamiltonian in short-hand notation as
\begin{equation}
\hat H = \sum_{\alpha\beta} h_{\alpha\beta} \hat S_{\alpha\beta} + \sum_{\alpha\beta\gamma\delta} h_{\alpha\beta\gamma\delta} \hat S_{\alpha\beta}\hat S_{\gamma\delta}.
\end{equation}
Here, $h_{\alpha\beta}$ are single-particle terms due to the magnetic field, and $h_{\alpha\beta\gamma\delta}$ are interaction terms. For our system, $h_{\alpha\beta}$ is a diagonal matrix. The Heisenberg equations for the multilevel operators can be written schematically as
\begin{equation}\label{eqn: eqns of motion}
i \partial_t \hat S_{\alpha\beta}(t) = [\hat S_{\alpha\beta}, \hat H] = \sum_{\mu\nu} \Lambda_{\alpha\beta}^{\mu\nu} \hat S_{\mu\nu} + \sum_{\mu\nu\lambda\sigma} \Lambda_{\alpha\beta}^{\mu\nu\lambda\sigma} \hat S_{\mu\nu}\hat S_{\lambda\sigma}
\end{equation}
where $\Lambda_{\alpha\beta}^{\mu\nu}$ depends on the single-particle terms, and $\Lambda_{\alpha\beta}^{\mu\nu\lambda\sigma}$ depends on the interactions. To obtain the classical equations of motion, we replace quantum operators with their classical counterparts. Monomial operators are replaced as $\hat S_{\alpha\beta} \to S_{\alpha\beta}$, and quadratic operators are replaced by a symmetric decoupling scheme, $\frac{1}{2} \{\hat S_{\alpha\beta}, \hat S_{\gamma\delta}\} \to S_{\alpha\beta}S_{\gamma\delta}$. Subsequently, we integrate these equations using a standard differential equation solver.

Observables are computed by averaging over trajectories. For instance, occupations are obtained from $n_\alpha(t) = \overline{ S_{\alpha\alpha}(t) }$, where $\overline{ \cdots }$ denotes ensemble average.

\begin{figure}[t] \centering
\includegraphics[width=0.7\columnwidth]{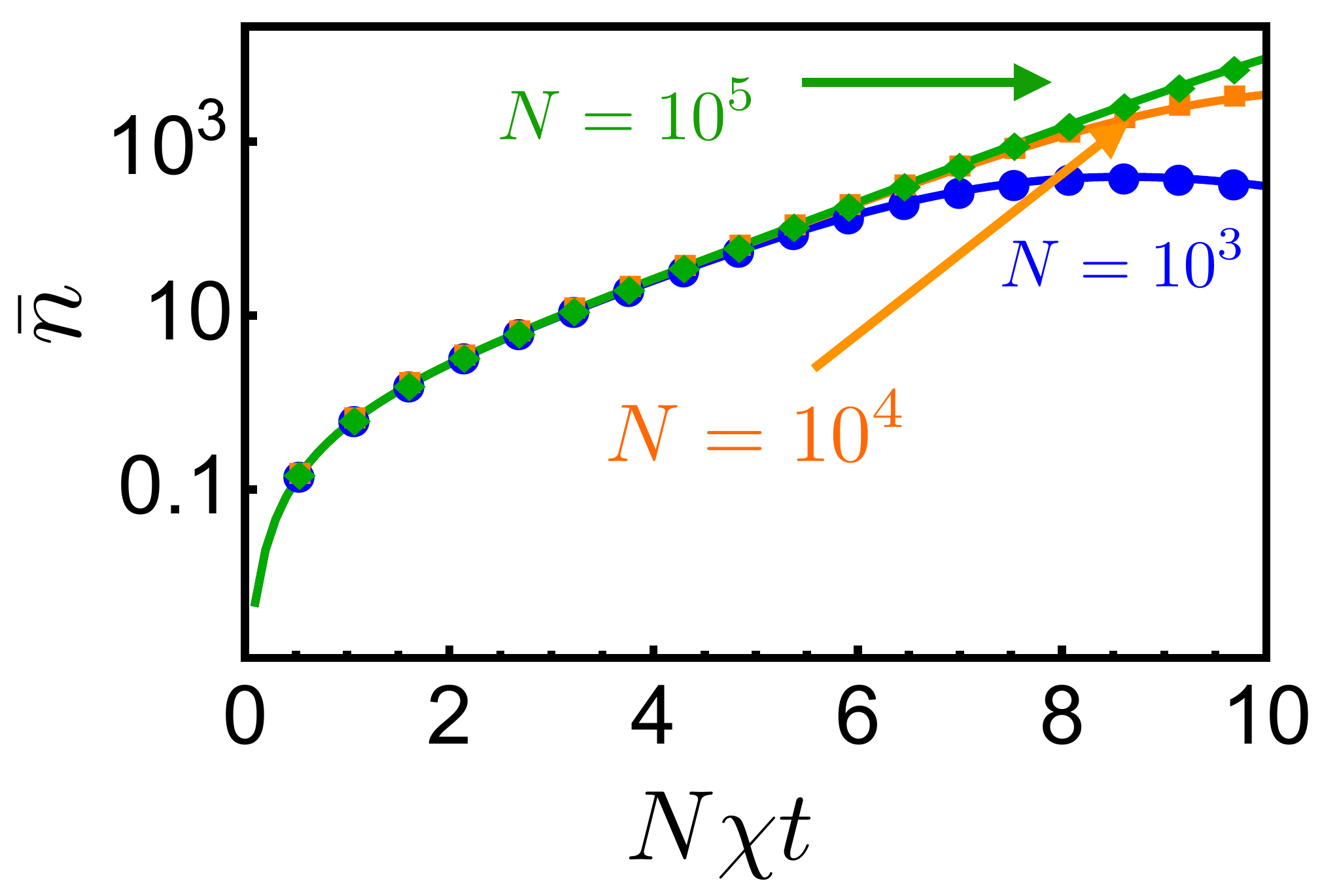}
\caption{Growth of number of entangled particles $\bar{n}$ for $\delta = N\chi/2$. Results obtained from TWA treatment of 20-level spin model (dots) agree excellently with results from numerical exact diagonalization (solid lines) of the reduced 4-level spin model [Eq.~(1) in the main text].}
\label{fig: supp population}
\end{figure}

As a first comparison, in Fig.~\ref{fig: supp population} we compute the number of entangled pairs, $\nbar$, that populate the levels $\vert g, F\rangle$ and $\vert e, -F\rangle$. We find excellent agreement between TWA results, using a model with 20 levels, and a numerically exact calculation for the four-level model.

\begin{figure}[t] \centering
\includegraphics[width=1.0\columnwidth]{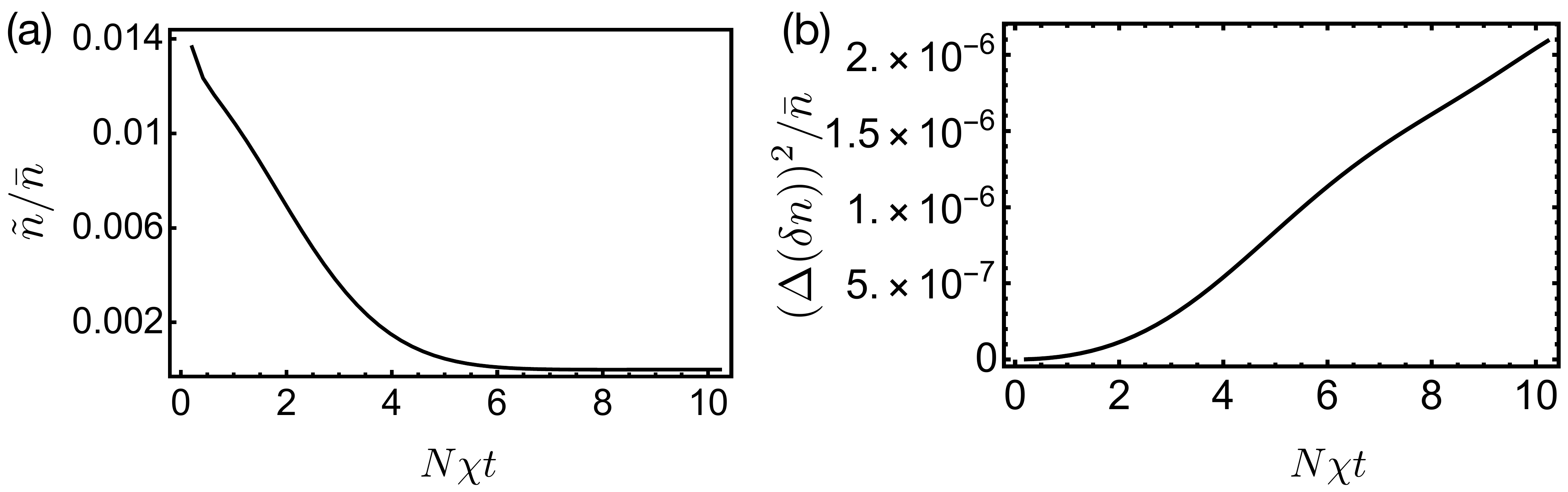}
\caption{
(a) Total occupation $\tilde{n}$ of the hyperfine levels $\ket{g,m\neq\pm F}$ and $\ket{e,m\neq\pm F}$, normalized to the occupation $\nbar$ in $\ket{g,F}$ and $\ket{e,-F}$. (b) Variance $(\Delta(\delta n))^2$ of the occupation difference $\delta n$ between $\ket{g,F}$ and $\ket{e,-F}$, normalized by $\nbar$. Both (a) and (b) are calculated using TWA for the full (20-level) spin model [Eq.~\eqref{eqn: H}] and using fixed $\delta = N\chi/2$ and $N=10^5$. Both $\tilde{n}$ and $(\Delta(\delta n))^2$ remain vanishingly small compared to $\nbar$ throughout the dynamics (note vertical scales).
}
\label{fig: supp twa2}
\end{figure}

To complement this result, in Fig.~\ref{fig: supp twa2} we use TWA to demonstrate that leakage of population outside the four-level model is negligible. 
Fig.~\ref{fig: supp twa2}(a) plots the total occupation $\tilde{n}$ in the hyperfine levels $\ket{g,m\neq\pm F}$ and $\ket{e,m\neq\pm F}$. This remains small and is relatively suppressed as time increases, validating our conjecture that the imbalance of the Clebsch-Gordoan coefficients will lead to a runaway dominance of the quartet of levels $\ket{g,\pm F}$ and $\ket{e,\pm F}$. Moreover, Fig.~\ref{fig: supp twa2}(b) plots the variance of the difference in population $\delta n$ in $\ket{e,-F}$ and $\ket{g,F}$. Vanishing fluctuations in the population difference are a hallmark of the bosonic two-mode squeezing that is predicted for our simplified four-level model (see later discussion of UPA), whereas the involvement of the other 16 levels would lead to the introduction of excess population fluctuations. Our results in Fig.~S2(b) demonstrate that these fluctuations remain small, relative to the level of uncorrelated Poissonian fluctuations $\sim \bar{n}$, throughout the timescales of interest for our protocol.

As a final subtle point, we note that the TWA method used in the treatment of the 20-level model was independently benchmarked by comparing a TWA treatment of the 4-level model to results generated by exact diagonalization. We found excellent agreement for all relevant variables across the timescales of interest in this proposal, which justifies our our use of TWA for the full 20-level model, for which direct comparison to numerical exact diagonalization is not feasible for meaningful system sizes.

\subsection{Validity of the UPA}
\begin{figure}[t] \centering
\includegraphics[width=0.7\columnwidth]{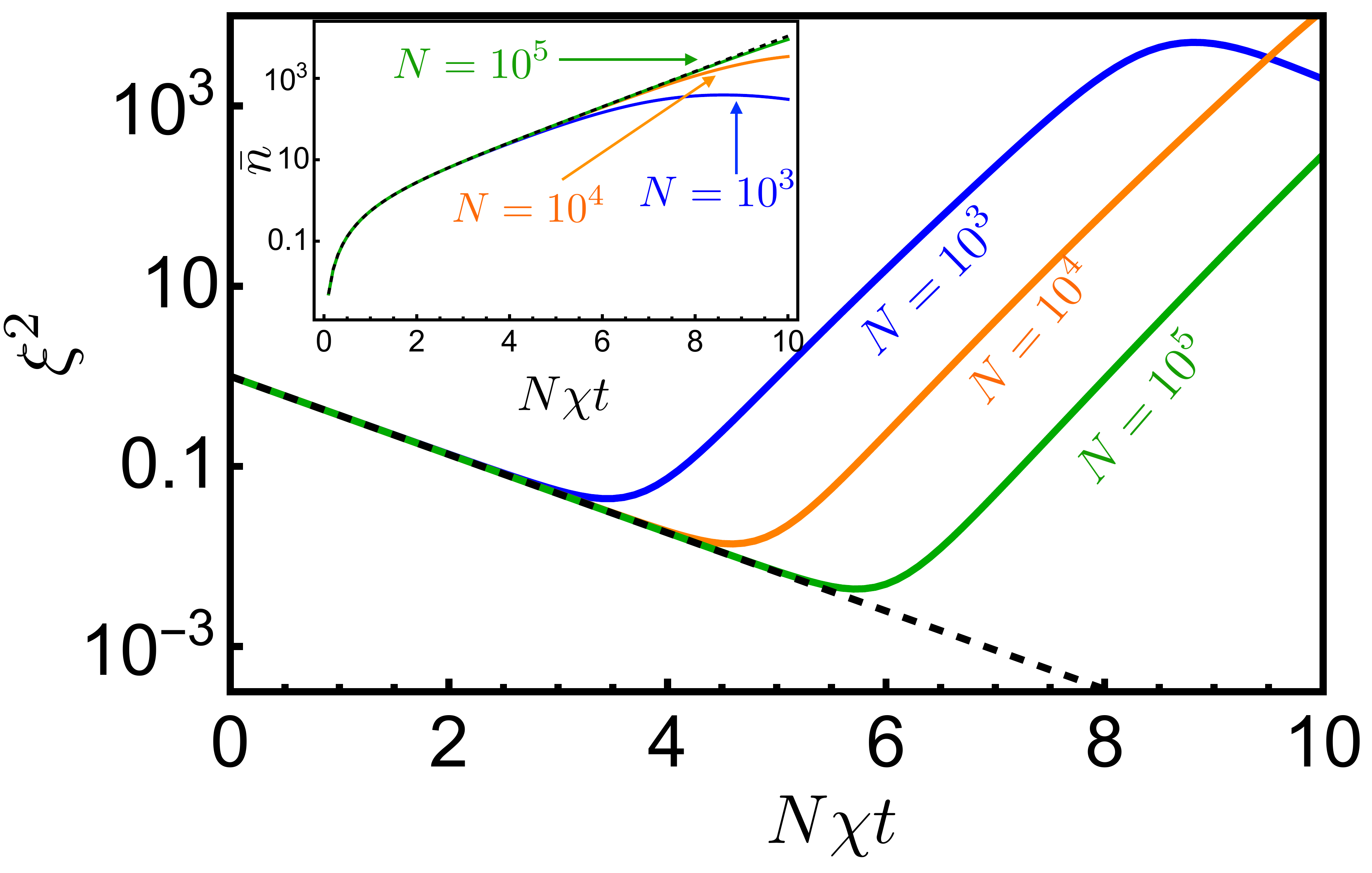}
\caption{
The squeezing $\xi^2$, calculated using numerical exact diagonalization of Eq.(1) in the main text with fixed $\delta = N\chi/2$. Inset: Number of entangled pairs $\nbar$ versus time.  Exact results agree with the UPA prediction (dashed lines) until finite-size effects (i.e., pump depletion) become important.}
\label{fig: ed}
\end{figure}
In the main text, we approximated the 4-level spin model [Eq.(1)] to a bosonic pair production model [Eq.~(3)] which is amenable to an analytic treatment when the UPA is implemented. We verify the UPA is applicable for short times in Fig.~\ref{fig: ed} by comparing to a numerically exact calculation for moderate system sizes ($N < 10^5)$. Specifically, we compute the number of entangled pairs $\nbar$ and the squeezing $\xi^2$ for fixed $\delta = N\chi/2$. The results in Fig.~\ref{fig: ed} show excellent agreement between the ED and UPA predictions at short timescales, as expected, before finite size effects (i.e, pump depletion) become relevant.

\section{Description of the squeezing in joint Bloch spheres}
An important and well known feature of the bosonic model of pair production [Eqs.~(\ref{eqn:Hbosonfull}) and (\ref{eqn: upa})] is that it generates squeezing in joint quadratures of the two bosonic modes. In this section, we show that this implies that, within UPA, our spin model is also capable of generating spin squeezing of combined spin quadratures of the $A$ and $B$ ensembles.

To be concrete, we consider only the case with $\delta = N\chi/2$ so that our effective spin Hamiltonian $\hat H$ [see Eq.~(1) of the main text)] reduces to 
\begin{equation} \label{eqn: pair production for spins}
\hat H = \hbar\chi(\hat S^+_A \hat S^-_B + \hat S^+_B \hat S^-_A).
\end{equation}
We begin by rewriting this Hamiltonian in a more convenient form, 
\begin{equation}
\hat H = 2\hbar\chi(\hat S^x_A \hat S^x_B + \hat S^y_A \hat S^y_B), 
\end{equation}
which can be alternatively factorized as,
\begin{multline} \label{eqn: factorization}
\hat H = \hbar\chi\left(\hat S^y_B + \hat S^x_A\right)\left(\hat S^x_B + \hat S^y_A\right) \\ - \hbar\chi\left(\hat S^y_B - \hat S^x_A\right)\left(\hat S^x_B - \hat S^y_A\right).
\end{multline}

To proceed, let us denote the terms appearing in Eq.~\eqref{eqn: factorization} as $\mathcal{S}_{1,+} = \hat S^x_B + \hat S^y_A$, $\mathcal{S}_{2,+} = \hat S^y_B + \hat S^x_A$, $\mathcal{S}_{1,-} = \hat S^x_B - \hat S^y_A$, and $\mathcal{S}_{2,-} = \hat S^y_B - \hat S^x_A$. As we will see below, this new notation will enable us to gain physical intuition by visualizing the physics as occurring on two independent Bloch spheres. We define a third axis, $\mathcal{S}_{3,-} = \hat S^z_B - \hat S^z_A$, 
which enables us to identify that $(\mathcal{S}_{1,+}, \mathcal{S}_{2,+}, \mathcal{S}_{3,-})$ and $(\mathcal{S}_{1,-}, \mathcal{S}_{2,-}, \mathcal{S}_{3,-})$ each independently satisfy the canonical commutation relations of SU(2) spins. Therefore, we can construct two Bloch spheres with a shared axis, with the first sphere's axes as $(\mathcal{S}_{1,+}, \mathcal{S}_{2,+}, \mathcal{S}_{3,-})$ and the second sphere's axes as $(\mathcal{S}_{1,-}, \mathcal{S}_{2,-}, \mathcal{S}_{3,-})$. The first line of Eq.~\eqref{eqn: factorization} acts only on the first Bloch sphere, and the second line acts only on the second Bloch sphere.

The generation of squeezing by $\hat H$ can be understood by mapping the axes of the Bloch spheres to bosonic quadratures, which translates $\hat H$ to a form that is familiar in quantum optics. At the mean-field level (equivalently, in the UPA), $\mathcal{S}_{3,-}$ is a constant of motion with value $N/2$. Therefore, in this approximation, $\mathcal{S}_{1,+}$ and $\mathcal{S}_{2,+}$ can be mapped to bosonic quadratures $\hat X_+$ and $\hat Y_+$ as $\mathcal{S}_{1,+} \approx \sqrt{\frac{N}{2}} \hat X_+$ and $\mathcal{S}_{2,+} \approx \sqrt{\frac{N}{2}} \hat Y_+$, which satisfy the canonical commutation relation $[\hat X_+, \hat Y_+] = i$. We similarly define $\mathcal{S}_{1,-} \approx \sqrt{\frac{N}{2}} \hat X_-$ and $\mathcal{S}_{2,-} \approx \sqrt{\frac{N}{2}} \hat Y_-$, where $[\hat X_-, \hat Y_-] = i$. The Hamiltonian, written in terms of these bosonic quadratures, is
\begin{equation}
\hat H = \frac{N\hbar\chi}{2}\hat Y_+\hat X_+ - \frac{N\hbar\chi}{2}\hat Y_-\hat X_-.
\end{equation}
It is well-known that this Hamiltonian generates squeezing along $\hat Y_+$ and $\hat X_-$ for $\chi > 0$ (and conversely, along $\hat X_+$ and $\hat Y_-$ for $\chi<0$) \cite{agarwal2012quantum}. Equivalently, we must have that our Hamiltonian generates squeezing in two \emph{spin} quadratures along $\mathcal{S}_{2,+}$ and $\mathcal{S}_{1,-}$ respectively for $\chi > 0$. This is the \textit{two-mode squeezing} that we report in the main text. The squeezing manifests itself as reduced quantum noise,
\begin{equation}
    \mathrm{var}(\mathcal{S}_{1,-}) = \mathrm{var}(\mathcal{S}_{2,+}) = \frac{N}{4}e^{-\chi N t},
\end{equation}
and the antisqueezing as increased quantum noise,
\begin{equation}
    \mathrm{var}(\mathcal{S}_{2,-}) = \mathrm{var}(\mathcal{S}_{1,+}) = \frac{N}{4}e^{\chi N t} .
\end{equation}
The squeezing and antisqueezing are well-defined relative to the isotropic projection noise associated with a typical coherent spin state on each of the independent collective Bloch spheres, characterized by $\mathrm{var}(\mathcal{S}_{1,\pm}) = \mathrm{var}(\mathcal{S}_{2,\pm}) = N/4$ for our initial state polarized along $\mathcal{S}_{3,-}$.

Lastly, we briefly discuss the illustrations of the quantum noise in Fig.~1(c) of the main text. The dynamics, and in particular the squeezing, produced by the bosonic Hamiltonian $\hat H = \frac{N\hbar\chi}{2}\hat Y_+\hat X_+ - \frac{N\hbar\chi}{2}\hat Y_-\hat X_-$ can be understood via the Wigner function computed in terms of the bosonic quadratures. It is well known that the Wigner function of two-mode squeezed vacuum factorizes into a pair of single-mode squeezed states with respect to the independent phase-spaces defined by $(X_+,Y_+)$ and $(X_-,Y_-)$. The axes describing these two phase spaces commute with each other. Strictly, when the UPA is satisfied we can transplant each of these bosonic phase-spaces (up to prefactors) to lie on the surface of the independent Bloch spheres defined by $(\mathcal{S}_{1,+}, \mathcal{S}_{2,+}, \mathcal{S}_{3,-})$ and $(\mathcal{S}_{1,-}, \mathcal{S}_{2,-}, \mathcal{S}_{3,-})$ by ignoring the curvature effects [i.e., the spin Wigner functions are effectively limited to the $2$D planes $(\mathcal{S}_{1,+}, \mathcal{S}_{2,+})$ and $(\mathcal{S}_{1,-}, \mathcal{S}_{2,-})$, where the spin Wigner functions on each Bloch sphere lie on axes that commute with their counterparts on the other Bloch sphere, assuming the pump populations are equal]. We adopt this correspondence for Fig.~1(c) to schematically illustrate the spin-squeezing in this spirit. For illustrative purposes, we enlarge the area covered by the spin Wigner functions, while strictly within the UPA, they are restricted to an infinitesimal surface perpendicular to $\mathcal{S}_{3,-}$.

\section{Evolution of the joint Bloch spheres during the Ramsey sequence}
\begin{figure}[t]\centering
\includegraphics[width=1.0\columnwidth]{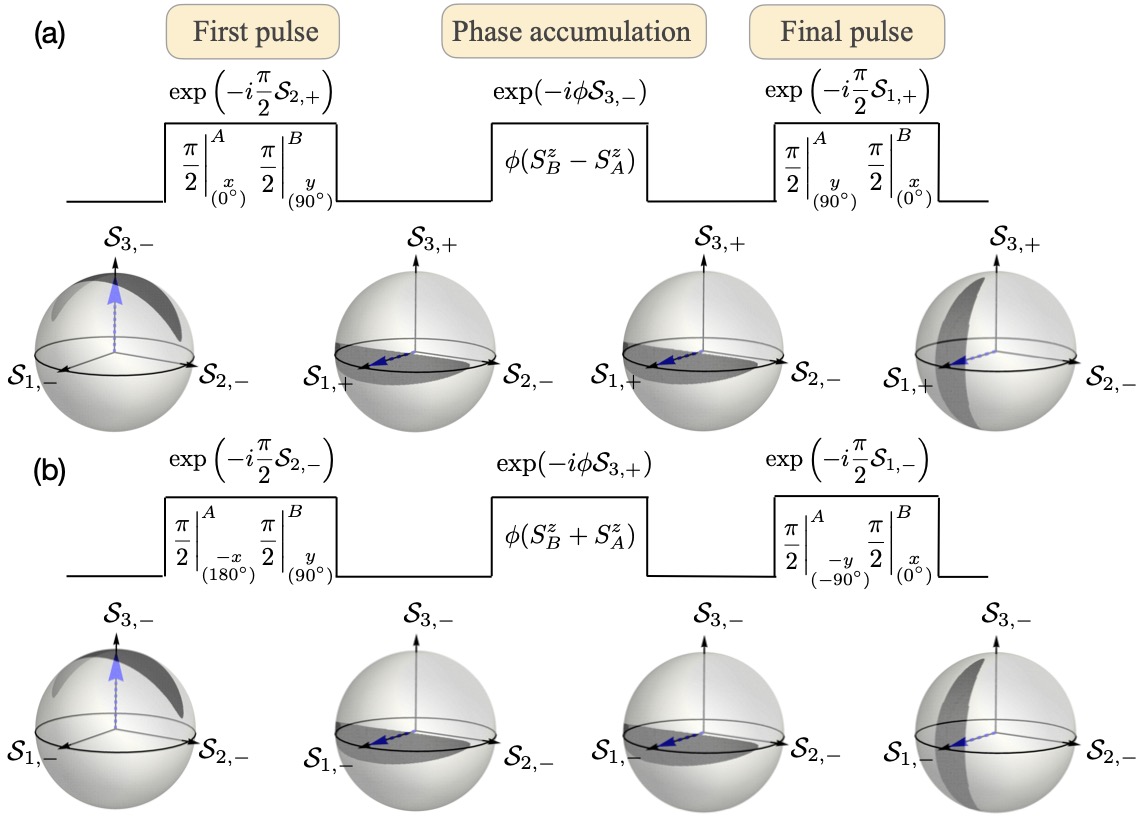}
\caption{Evolution of the spin distributions in the upper Bloch sphere in Fig.~1(c) in the main text, during the Ramsey sequence for measuring (a) the differential phase, (b) the sum phase. Information about the imprinted phase is not captured in the spheres shown here. Dashed blue arrows show the Bloch vector.}
\label{suppfig: other Bloch sphere}
\end{figure}

\begin{figure}[t]\centering
\includegraphics[width=1.0\columnwidth]{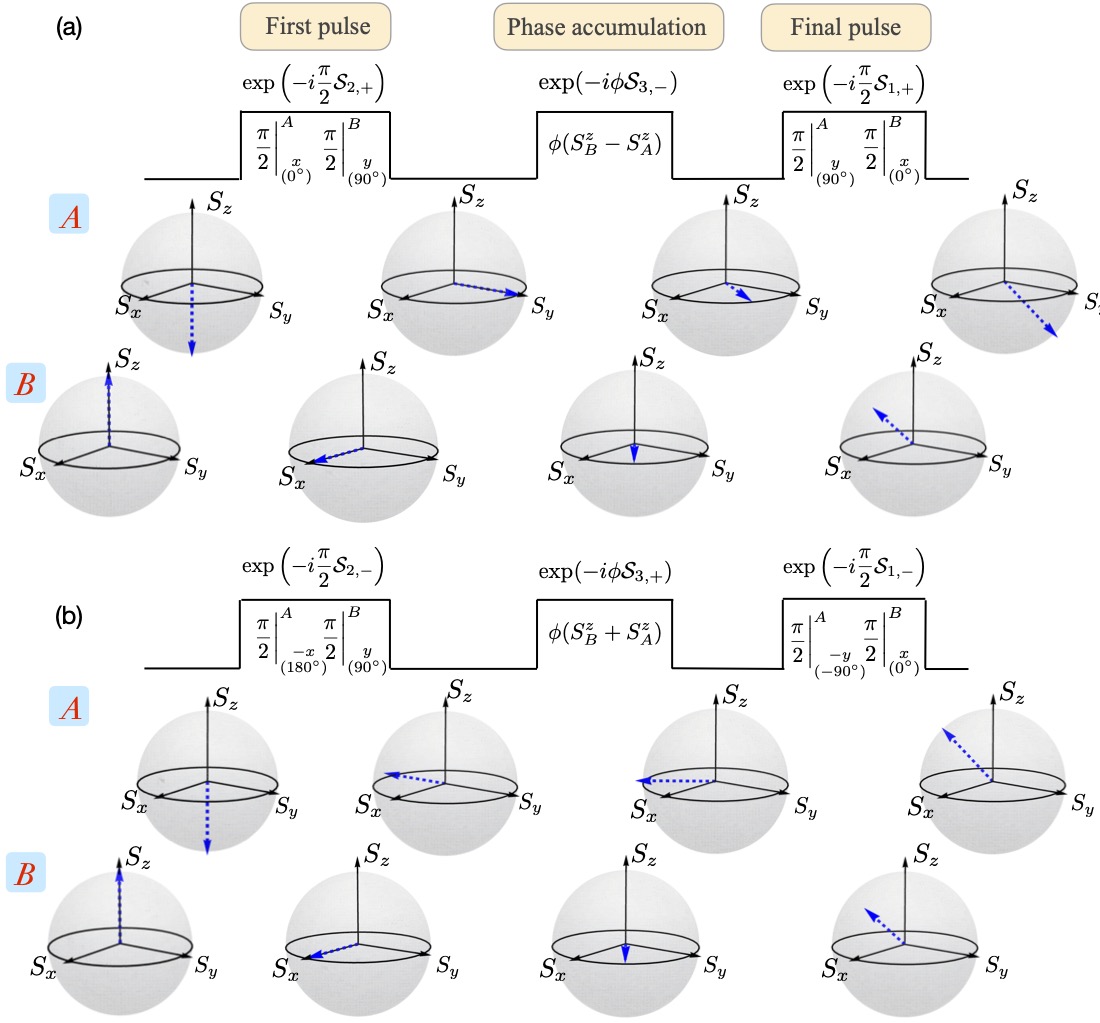}
\caption{Rotation of the Bloch vectors in the $A$ and $B$ ensembles during the Ramsey sequence for measuring (a) the differential phase, (b) the sum phase.}
\label{suppfig: Bloch vectors}
\end{figure}

In the main text, we visualized the Ramsey sequence as rotations of the spin distribution on the lower Bloch sphere of Fig.~1(c). Here, we describe the evolution of the spin distribution on the upper Bloch sphere during the Ramsey sequence.

We first consider the Ramsey sequence that measures the differential phase [illustrated in Fig.~2(a) in the main text]. After the first pulse, realized by $\exp\left(-i\frac{\pi}{2}\hat{\mathcal{S}}_{2,+}\right)$, the axes on the upper Bloch sphere are rotated from $(\mathcal{S}_{1,-}, \mathcal{S}_{2,-}, \mathcal{S}_{3,-})$ to $(-\mathcal{S}_{3,+}, \mathcal{S}_{2,-}, \mathcal{S}_{1,+})$. This is illustrated in Fig.~\ref{suppfig: other Bloch sphere}(a). The Bloch vector points along $\mathcal{S}_{1,+}$, which is now a shared axis with the other Bloch sphere [see Fig.~2(a)]. At this step, the system is a null state of $\mathcal{S}_{1,-}$, assuming the initial pump populations were equal.

The next step in the Ramsey sequence is a small rotation $\phi$ of the Bloch vector. In the main text, we considered a rotation generated by $\exp(-i\phi \mathcal{S}_{3,-})$. This operation does not rotate the spin distribution in Fig.~\ref{suppfig: other Bloch sphere}(a). This is because in UPA, the spin distribution lies entirely in the plane $(\mathcal{S}_{3,+}, \mathcal{S}_{2,-})$, and these two axes commute with $\exp(-i\phi \mathcal{S}_{3,-})$, if the initial pump populations were equal. The rotation generated by $\exp(-i\phi \mathcal{S}_{3,-})$ only rotates the spin distribution in the Bloch sphere shown in the main text.

The final step in the Ramsey sequence is a rotation implemented by $\exp\left(-i\frac{\pi}{2}\mathcal{S}_{1,+}\right)$, which rotates the spin distribution by $90^\circ$. Figure \ref{suppfig: other Bloch sphere}(a) shows that the total atomic inversion encodes no information about $\phi$, i.e. $\braket{\hat S^z_A + \hat S^z_B} = 0$. Conversely, Fig.~2(a) (main text)  showed that the difference in atomic inversions, $\braket{\hat S^z_B - \hat S^z_A}$, encoded information about $\phi$.

Fig.~\ref{suppfig: other Bloch sphere}(b) shows how the upper Bloch sphere in Fig.~1(c) evolves during the Ramsey protocol that measures the sum phase. No information about the imprinted sum phase is captured by the sphere in Fig.~\ref{suppfig: other Bloch sphere}(b).

For ease of understanding, we also show in Fig.~\ref{suppfig: Bloch vectors} how the Bloch vectors in the $A$ and $B$ ensembles separately evolve during the two Ramsey protocols.

\section{Achievable sensitivity incorporating intrinsic decoherence}
In this section, we sketch how we compute the sensitivity of our protocol in the presence of decoherence. We begin by writing the master equation for $\hat S_A^\pm$ and $\hat S_B^\pm$, analogous to Eq.~\eqref{eqn: heisenberg eqn unitary}, but now including decoherence as well. The two sources of decoherence that we consider are collective decay and independent spontaneous decay of atoms from the excited states. These two decoherences have two main effects: they reduce coherences, and cause depletion of the excited modes. This will require us to go beyond the UPA, since the excited mode in the $B$ ensemble is a pump mode, and it gets depleted by decay to the ground state. As a further effect of pump depletion, the Zeeman shift does not cancel the mean-field energy shift. We will account for this as well.

The master equation is
\begin{align} \label{eqn: master eqn}
\partial_t \left( \begin{array}{c} \braket{ \hat S_A^-} \\ \braket{\hat S_B^-} \end{array} \right) \approx &
\left( \begin{array}{cc} -\frac{\gamma}{2} + i\delta & 0 \\ 0 & -\frac{\gamma}{2} - i\delta \end{array} \right)  \left( \begin{array}{c} \braket{ \hat S_A^-} \\ \braket{\hat S_B^-} \end{array} \right) \nonumber\\
 &+ (\Gamma+2i\chi) \left( \begin{array}{cc} \braket{\hat S_A^z} & \braket{\hat S_A^z} \\ \braket{\hat S_B^z} & \braket{\hat S_B^z} \end{array} \right) \left( \begin{array}{c} \braket{ \hat S_A^-} \\ \braket{\hat S_B^-} \end{array} \right),
\end{align}
where we have approximated that $\braket{ \hat S_A^z \hat S_B^-} \approx \braket{ \hat S_A^z }\braket{ \hat S_B^-}$ and similarly for $A \leftrightarrow B$. Since spontaneous decay causes the excited populations to decrease with time, we have to modify the undepleted pump approximation, $\braket{ \hat S_B^z} = \frac{N}{4}$, to $\braket{ \hat S_B^z} = \frac{N}{4}(2e^{-\gamma t}-1)$. We set $\braket{ \hat S_A^z} = -N/4$, since there is no initial excited population in the $A$ ensemble. We choose $\delta = \frac{N\chi}{2}$ as we did in the main text.

Equation~\eqref{eqn: master eqn} is a time-dependent coupled differential equation which is non-trivial to solve. Instead, we solve Eq.~\eqref{eqn: master eqn} to first order in $\Gamma$ and $\gamma$. Separating the matrix on the right hand side as $M = M_0 + M_1(t)$, where $M_0$ is the contribution at $\Gamma = \gamma=0$, and $M_1(t)$ is the remaining contribution, the solution to Eq.~\eqref{eqn: master eqn} is
\begin{align}\label{eqn: master eqn soln}
\left( \begin{array}{c} \braket{ \hat S_A^-(t)} \\ \braket{\hat S_B^-(t)} \end{array} \right) = & \left(e^{M_0t} + \int_0^t d\tau\ e^{M_0(t-\tau)} M_1(\tau) e^{M_0\tau}\right) \nonumber\\
&\times \left( \begin{array}{c} \braket{ \hat S_A^-(0)} \\ \braket{\hat S_B^-(0)} \end{array} \right).
\end{align}
Similar equations can be set up and solved for the second moments, $\braket{\hat S_A^+ \hat S_B^-}$, $\braket{\hat S_A^+ \hat S_A^-}$, and $\braket{\hat S_B^+ \hat S_B^-}$, which are necessary for calculating the squeezed quantum noise.

In the Ramsey protocol that measures the differential phase, the value of the signal measured is
\begin{align}\label{eqn: signal with decoherence}
\braket{ \hat S_B^z - \hat S_A^z}_{\rm final} =& \braket{ \hat S_B^z(t) - \hat S_A^z(t)}\sin\phi \nonumber\\ &+ \braket{ \hat S_B^y(t) + \hat S_A^x(t) }\cos\phi,
\end{align}
where $t$ is the time before the first $\pi/2$ pulse in the Ramsey sequence. The first term in the above equation is $-\frac{N}{2}e^{-\gamma t}\sin\phi$, and the second term is zero. The fluctuation in the signal is
\begin{align}\label{eqn: variance with decoherence}
&( \Delta(\hat S_B^z - \hat S_A^z))^2_{\rm final} =  (\Delta(\hat S_B^z(t) - \hat S_A^z(t)))^2\sin^2\phi \nonumber\\
&+ ( \Delta(\hat S_B^y(t) + \hat S_A^x(t)))^2 \cos^2\phi\nonumber\\
&+ \cos\phi\sin\phi \braket{ (\hat S_B^z(t) - \hat S_A^z(t)) (\hat S_B^y(t) + \hat S_A^x(t))} \nonumber\\
&+ \cos\phi\sin\phi \braket{ (\hat S_B^y(t) + \hat S_A^x(t)) (\hat S_B^z(t) - \hat S_A^z(t)) }.
\end{align}
The best sensitivity is achieved at $\phi = 0$, therefore we set $\phi = 0$ hereafter. Solving for the second order moments in a similar fashion to Eq.~\eqref{eqn: master eqn soln}, the fluctuation in the signal is
\begin{align}\label{eqn: variance with decoherence}
& ( \Delta(\hat S_B^z - \hat S_A^z))^2_{\rm final} = \frac{N}{4} \left( e^{-N\chi t} + \frac{\Gamma}{2\chi} + \frac{\gamma}{N\chi}  \right. \nonumber\\
& \left. + e^{N\chi t}\left(\frac{\Gamma}{4\chi} + \frac{\gamma}{2N\chi} - \frac{\gamma t}{2}\right)^2  + e^{N\chi t}\left(\frac{\gamma}{2N\chi} - \frac{\gamma t}{2}\right)^2 \right).
\end{align}
This yields the sensitivity as
\begin{align}
(\Delta\phi)^2 = & \frac{e^{-N\chi t}}{N} + \frac{\Gamma}{2N\chi} + \frac{\gamma}{N^2\chi} + \frac{e^{N\chi t}}{N} \left(\frac{\gamma}{2N\chi} - \frac{\gamma t}{2}\right)^2 \nonumber\\
& + \frac{e^{N\chi t}}{N} \left(\frac{\Gamma}{4\chi} + \frac{\gamma}{2N\chi} - \frac{\gamma t}{2}\right)^2.
\end{align}

An analysis for the Ramsey protocol that measures the sum phase yields the same expression for the sensitivity.

\section{Scaling of $(\Delta\phi_{\rm min})^2$ with $N$.}
The best sensitivity is achieved at optimum values of $\Delta$ and $t$. These optimum values aim to find the right balance between the gain obtained by reaching a higher $\nbar$ versus the loss in squeezing due to decoherence. We find the optimum values of $\Delta$ and $t$ by setting the derivatives of $\Delta\phi^2$ with respect to $\Delta$ and $t$ as zero.

First, we find the optimum duration for our protocol. This is obtained by setting $d(\Delta\phi)^2/dt = 0$, which yields the implicit equation
\begin{equation}
\gamma t = \frac{\Gamma}{4\chi} + \sqrt{2e^{-2N\chi t} + \left(\frac{\gamma}{N\chi}\right)^2 - \left(\frac{\Gamma}{4\chi}\right)^2}.
\end{equation}
We approximate this time as $\gamma t \approx \frac{\Gamma}{4\chi} + \sqrt{2}e^{-N\Gamma/4\gamma}$. At this optimum time, the sensitivity is given by
\begin{align}
N(\Delta\phi)^2 = & \exp\left(-\frac{N\Gamma}{4\gamma} - \frac{\sqrt{2}N\chi}{\gamma}e^{-N\Gamma/4\gamma} \right) + \frac{\Gamma}{2\chi} + \frac{\gamma}{N\chi} \nonumber\\
&+ \exp\left(\frac{N\Gamma}{4\gamma} + \frac{\sqrt{2}N\chi}{\gamma}e^{-N\Gamma/4\gamma} \right)\nonumber\\
&\times \left( \frac{\gamma^2}{N^2\chi^2} + e^{-N\Gamma/2\gamma} - \frac{\sqrt{2}\gamma}{N\chi}e^{-N\Gamma/4\gamma} \right).
\end{align}
The parameters $\Gamma$ and $\chi$ depend on the cavity detuning $\Delta$ and the cavity loss rate $\kappa$ as $\chi = \frac{g_F^2}{\Delta}$ and $\Gamma = \frac{g_F^2\kappa}{\Delta^2}$. Rewriting the sensitivity in terms of $\Delta$, $\kappa$, and the cavity cooperativity $C = \frac{4g_F^2}{\kappa\gamma}$, we optimize the sensitivity with respect to $\Delta$. This gives us an optimum value of $\Delta$,
\begin{align}
\Delta \approx \sqrt{\frac{NC}{2\ln(2NC)}} \kappa.
\end{align}
The optimum sensitivity at this detuning is given by
\begin{equation}
N(\Delta\phi)^2 \approx \sqrt{\frac{2\ln(2NC)}{NC}}.
\end{equation}

\section{Effect of pump population fluctuations on the sensitivity}\label{sec: number fluctuation}
Here, we calculate the sensitivity when the pump populations differ from the idealized case of $N/2$. The populations of the pump states $\ket{g,A}$ and $\ket{e,B}$ are parameterized as $N_{g,A} = (N + \delta N_{\rm tot} + \delta N_{AB})/2$ and $N_{e,B} = (N + \delta N_{\rm tot} - \delta N_{AB})/2$, where $\delta N_{AB}$ and $\delta N_{\rm tot}$ characterize fluctuations of the difference and sum of the populations of the $A$ and $B$ ensembles, respectively.

The expression of the relevant spin variables in terms of the Schwinger bosons $\hat a_{g,B}$ and $\hat a_{e,A}$ is then,
\begin{align}
&\mathcal{S}_{1,\pm} \approx \sqrt{N_{e,B}}\frac{\hat a_{g,B}^{\phantom\dagger} + \hat a_{g,B}\+}{2} \pm \sqrt{N_{g,A}}\frac{\hat a_{e,A}\+ - \hat a_{e,A}^{\phantom\dagger}}{2i}, \nonumber\\
&\mathcal{S}_{2,\pm} \approx \sqrt{N_{e,B}}\frac{\hat a_{g,B}^{\phantom\dagger} - \hat a_{g,B}\+}{2i} \pm \sqrt{N_{g,A}}\frac{\hat a_{e,A}\+ + \hat a_{e,A}^{\phantom\dagger}}{2},\nonumber\\
&\mathcal{S}_{3,-} = \frac{N_{e,B} + N_{g,A}}{2} - \hat a_{e,A}\+\hat a_{e,A}^{\phantom\dagger} - \hat a_{g,B}\+\hat a_{g,B}^{\phantom\dagger} \approx \frac{N + \delta N_{\rm tot}}{2}.
\end{align}
From these relations, we obtain
the Heisenberg equation of motion for $\ha_{e,A}$ and $\ha_{g,B}$ within UPA as
\begin{equation}\label{eqn: eom for different pump population}
i\partial_t \left( \begin{array}{c} \hat a_{e,A} \\ \hat a_{g,B}\+ \end{array} \right) = 
\left( \begin{array}{cc} N_{g,A}\chi - \delta & \sqrt{N_{g,A} N_{e,B}}\chi \\ -\sqrt{N_{g,A} N_{e,B}}\chi & -N_{e,B}\chi + \delta \end{array} \right)
\left( \begin{array}{c} \hat a_{e,A} \\ \hat a_{g,B}\+ \end{array} \right).
\end{equation}
These are solved analytically to obtain relevant expectation values, as per the ideal case previously discussed. 

We focus on the effect of population fluctuations on the sensitivity of our Ramsey protocol, which for $\phi = 0$ is given by
\begin{equation}
(\Delta\phi)^2 = \frac{ {\rm var}(\mathcal{S}_{1,-}(t)) }{ \braket{\mathcal S_{3,-}(t)}^2 }.
\end{equation}
Here, we have used that the relevant squeezed quadrature remains $\mathcal{S}_{1,-}$ even in the presence of small population fluctuations (which can be explicitly shown by diagonalizing the associated covariance matrix of the squeezed state). 

Assuming that the sum and difference of the initial pump populations vary between shots of an experiment as a Gaussian random variable with zero mean and rms fluctuations $\sigma_{AB}$ and $\sigma_{\mathrm{tot}}$, respectively, the relevant expectation values generated by the solution of Eq.~\eqref{eqn: eom for different pump population} are
\begin{align}
\bar{{\rm var}(\mathcal{S}_{1,-})} \approx &  \frac{N}{4}e^{-N\chi t} + \frac{\sigma_{\rm tot}^2 + \sigma_{AB}^2}{16N}e^{N\chi t}, \nonumber\\
\bar{\braket{\mathcal{S}_{3,-}}} \approx \frac{N}{2}.
\end{align}
The approximate sign indicates that these results have been generated under the assumption that $\sigma_{AB}, \sigma_{\rm tot} \ll N$.
The sensitivity is thus
\begin{equation}\label{eqn:sens_Nfluct}
(\Delta\phi)^2 \approx \frac{e^{-N\chi t}}{N} + \frac{\sigma_{\rm tot}^2 + \sigma_{AB}^2}{4N^3}e^{N\chi t}.
\end{equation}
Insight into this result can be obtained by recalling that finite size effects beyond UPA limit the achievable squeezing to $(\Delta\phi)^2 \sim N^{-3/2}$ and reached at timescales $e^{N\chi t}\sim \sqrt{N}$ (note that decoherence limits the sensitivity similarly). Making the replacement $e^{N\chi t} \to \sqrt{N}$ in the latter terms of Eq.~(\ref{eqn:sens_Nfluct}) we thus determine that if the population fluctuations are sufficiently small, i.e., if $\sigma_{AB}, \sigma_{\mathrm{tot}} \lesssim \sqrt{N}$, then the corrections to the sensitivity will also be $O(N^{-3/2})$. Phrased alternatively, our predicted sensitivity is robust to number fluctuations at the level of the shot noise.

\section{Pair production with spin-1 BECs}
In the main text we contrast our effective bosonic model with alternative realization in quantum optics and spinor Bose-Einstein condensates. In the latter case, degenerate four-wave mixing is engineered through $s$-wave atomic collisions that change the internal Zeeman state of the atoms \cite{law1998}. It is common to simplify the theoretical treatment of the atomic collisions by invoking a single mode approximation (SMA), wherein it is assumed that all atoms share the same spatial wavefunction, regardless of their internal Zeeman state. This is typically valid for small systems and short to intermediate timescales, although a range of factors contribute \cite{jie2020}. 

By freezing out the spatial degree of freedom via the SMA the internal state dynamics for a spin-$1$ BEC is then given by, 
\begin{equation}\label{eqn:HBEC}
\hat H_{\rm BEC} = \frac{U_s}{2N}\hat{\vec{S}} \cdot \hat{\vec{S}} + q(\hat n_1 + \hat n_{-1}).
\end{equation}
Here, the spin operators are $\hat S_x = (\hat a_1\+ \hat a_0^{\phantom\dagger} + \hat a_0\+ \hat a_{-1}^{\phantom\dagger} + {\rm h.c.})/\sqrt{2}$, $\hat S_y = (\hat a_1\+ \hat a_0^{\phantom\dagger} + \hat a_0\+ \hat a_{-1}^{\phantom\dagger} - {\rm h.c.})/\sqrt{2}i$, and $\hat S_z = \hat a_1\+ \hat a_1^{\phantom\dagger} - \hat a_{-1}\+ \hat a_{-1}^{\phantom\dagger}$, where $\hat a_m\+$ creates a boson in Zeeman sublevel $m=0,\pm1$. The first term of Eq.~(\ref{eqn:HBEC}) describes spin-mixing due to $s$-wave atomic collisions characterized by interaction strength $U_s$, whereas the second term describes a quadratic Zeeman shift $q \propto B^2$ due to an applied magnetic field. 

The Hamiltonian (\ref{eqn:HBEC}) can be rewritten in a more insightful form, $\hat H_{\rm BEC} = \hat{H}_{\mathrm{inel}} + \hat{H}_{\mathrm{el}} + \hat{H}_{\mathrm{Z}}$ with \cite{rls_su2_2021}
\begin{equation}
    \begin{gathered}
        \hat{H}_{\mathrm{inel}} = \frac{U_s}{N}\left( \hat{a}_0\hat{a}_0 \hat{a}^{\dagger}_1\hat{a}^{\dagger}_{-1} + h.c. \right) , \\
        \hat{H}_{\mathrm{el}} = \frac{U_s}{N} \hat{n}_0 \left( \hat{n}_1 + \hat{n}_{-1} \right) + \frac{U_s}{2N} \left( \hat{n}_1 - \hat{n}_{-1} \right)^2 , \\
        \hat{H}_{\mathrm{Z}} = q(\hat{n}_1 + \hat{n}_{-1}) .
    \end{gathered}
\end{equation}
The first term, $\hat{H}_{\mathrm{inel}}$, describes spin-changing collisions wherein a pair of $m=0$ atoms collide and scatter into an $m=\pm1$ pair. Conversely, the second term, $\hat{H}_{\mathrm{el}}$, describes elastic collisions which preserve the relative spin populations. Inspection of $\hat H_{\rm BEC}$ in this form shows that, up to the distinguishing feature that the spin-$1$ BEC involves only three bosonic modes, it is analogous to the four-wave mixing Hamiltonian for our system described in Eq.~(\ref{eqn:Hbosonfull}).

Many experiments probe the regime where a BEC is prepared with the vast majority of atoms in the $m=0$ state. For large systems the UPA then corresponds to replacing $\hat a_0 \approx \sqrt{N}$ where $N$ is the number of atoms in the condensate. Equation (\ref{eqn:HBEC}) then reduces to 
\begin{equation}
\hat H_{\rm BEC} \approx (q+U_s) (\hat a_1\+ \hat a_1^{\phantom\dagger} + \hat a_{-1}\+ \hat a_{-1}^{\phantom\dagger}) + U_s(\hat a_1\+ \hat a_{-1}\+ + {\rm h.c.}).
\end{equation}
where we have ignored a term $\propto \hat{a}^{\dagger}_1\hat{a}_1 - \hat{a}^{\dagger}_{-1}\hat{a}_{-1}$ as a conserved quantity. Within the UPA, the spin-$1$ Hamiltonian is thus analogous to $\hat H_{\rm TMS}$ in Eq.~(3) in the main text, with $U_s = N\hbar\chi/2$, and $q = \hbar\delta$. The resonant condition for pair production is met when $q = -U_s$.

\bibliography{refs}